\documentclass[fleqn,10pt]{wlscirep}
\usepackage[utf8]{inputenc}
\usepackage[T1]{fontenc}
\usepackage{subcaption}
\usepackage{algorithm}
\usepackage{algorithmic}
\title{Social Opinions Prediction Utilizes Fusing Dynamics Equation with LLM-based Agents}

\author[1]{Junchi Yao}
\author[1,*]{Hongjie Zhang}
\author[2]{Jie Ou}
\author[1]{Dingyi Zuo}
\author[3]{Zheng Yang}
\author[4]{Zhicheng Dong}
\affil[1]{Sichuan Normal University, College of Computer Science, Chengdu, China}
\affil[2]{University of Electronic Science and Technology of China, School of Information and Software Engineering, Chengdu, China}
\affil[3]{Fujian Normal University, Fujian Provincial Engineering Technology Research
Center of Photoelectric Sensing Application, Fuzhou, China}
\affil[4]{Tibet University, College of Information
Science Technology, Lhasa, China}
\affil[*]{zhanghongjie@sicnu.edu.cn}

\begin{abstract}
In the context where social media emerges as a pivotal platform for social movements and shaping public opinion, accurately simulating and predicting the dynamics of user opinions is of significant importance. Such insights are vital for understanding social phenomena, informing policy decisions, and guiding public opinion. Unfortunately, traditional algorithms based on idealized models and disregarding social data often fail to capture the complexity and nuance of real-world social interactions. This study proposes the Fusing Dynamics Equation-Large Language Model (FDE-LLM) algorithm. This innovative approach aligns the actions and evolution of opinions in Large Language Models (LLMs) with the real-world data on social networks. The FDE-LLM devides users into two roles: opinion leaders and followers. Opinion leaders use LLM for role-playing and employ Cellular Automata(CA) to constrain opinion changes. In contrast, opinion followers are integrated into a dynamic system that combines the CA model with the Susceptible-Infectious-Recovered (SIR) model. This innovative design significantly improves the accuracy of the simulation. Our experiments utilized four real-world datasets from Weibo. The result demonstrates that the FDE-LLM  significantly outperforms traditional Agent-Based Modeling (ABM) algorithms and LLM-based algorithms. Additionally, our algorithm accurately simulates the decay and recovery of opinions over time, underscoring LLMs potential to revolutionize the understanding of social media dynamics.
\end{abstract}
\begin{document}

\flushbottom
\maketitle
\thispagestyle{empty}

\section*{Introduction}

The dynamic of user opinions plays a crucial role in leading social movements and influencing public opinion. Predicting these dynamics is essential for understanding the social phenomena, scientifically formulating policies, and effectively steering  public discourse \cite{1pennycook2021shifting,2ginossar2022cross,3budak2011limiting,4noorazar2020recent}. The process of user opinion formation and evolution is complex and driven by multiple factors. Therefore, capturing this process accurately has become the key point of attention in both academic and industrial sectors.

Opinion simulation methods are divided  into two categories: traditional Agent-Based Modeling (ABM) and Large Language Model (LLM)-based algorithms. ABM models the interactions between individuals and local rules to simulate opinion diffusion. Cellular Automata (CA), with their structured grid dynamics and local spatial interactions, provide a powerful framework  to simulate the intricate spatiotemporal evolution of complex systems\cite{Hu2018CellularA,5Degroot1974ReachingAC}. Bounded confidence models integrate psychological factors into their frameworks, including the Deffuant-Weisbuch (DW) model \cite{6deffuant2000mixing} and the Hegselmann-Krause (HK) model \cite{7rainer2002opinion}. The voter model developed by Clifford \cite{8clifford1973model} and Holley \cite{9holley1975ergodic} captures the essence of the public choice dynamic. Furthermore, empirical ABMs are developed based on real data. Carpentras utilized ABM to simulate the phenomenon of attitude polarization based on Real Life Experiments (ABM-RLE) \cite{10carpentras2022deriving}, while Duggins proposed the Psychologically-Motivated Model 
 (ABM-PM) and reproduced two empirical datasets on Americans’ political opinions \cite{11duggins2014psychologically}. However, these models still can't resolve the large variance and self-decay problems when simulating the real world.

The LLM-based opinion propagation algorithm\cite{12tornberg2023simulating} offers an innovative approach using LLM agents to replace human social interactions to build a social network\cite{13park2022social,14chang2024llms, 15papachristou2024network}. This framework involves creating agents that engage in interactions to disseminate and predict opinion.  G. De Marzo demonstrated that agents in their interactions can spontaneously form scale-free networks, which resemble real-world social networks such as Twitter\cite{16de2023emergence}.  Additionally, researchers at Stanford have expanded LLMs' functionality to store memories and implement dynamical action planning \cite{17park2023generative}. Chuang et al. have identified inherent biases in LLM-based agents, aligning them with real-world scientific consensus \cite{18chuang2023simulating}. Furthermore, Liu et al. introduced a framework for an LLM-based fake news propagation simulation framework (FPS) \cite{19liu2024skepticism}. 

In the context of group simulation,  G. De Marzo defined a ``majority force coefficient'' to measure the tendency of agents to follow the majority opinion. They found that a LLM society can only spontaneously reach consensus if the group size does not exceed a certain threshold\cite{20de2024language}.
 To control the number of agents, Mou et al. have proposed a hybrid framework where core users are driven by LLMs while stimulating regular users using a deductive agent-based model. Nonetheless, this approach still struggles with predicting opinions on real-world events due to the unconstrained opinion diffusion \cite{21mou2024unveiling}.

Traditional opinion dynamics models, like CA and HK, fall short in simulating human boredom or immunity emotions. Repeated event propagation leads individuals to boredom, leading their opinions towards neutrality. At the same time, the epidemic model, notably the Susceptible-Infectious-Recovered (SIR) model, effectively illustrates how information spreads \cite{22cooper2020sir}. Therefore, human boredom can be modeled using the concept of SIR model.

This study introduces an innovative opinion dynamics simulation method called Fusing Dynamics Equation LLM (FDE-LLM), which integrates LLMs with the SIR model. Our approach designs a Weibo (\url{https://weibo.com/}) simulator, where opinion leaders use LLM to interact each other, while opinion followers use ABM for opinion diffusion. Then, we constrain the opinion changes of opinion followers by integrating CA with the SIR model, while the CA model constrains the opinion leaders. Finally, we employ few-shot prompts to guide the opinion leaders in outputting actions based on their opinion values, determined by CA and LLM reflections. Utilizing the open-source model ChatGLM as the LLM Agent, we perform experiments on four real Weibo datasets. The results demonstrate that our FDE-LLM significantly surpasses existing ABM and LLM-based algorithms on Dynamic Time Warping (DTW) distance and Pearson correlation coefficient(Corr.) metrics.

Our contributions are threefold:
\begin{itemize}
    \item We present the LLM-based opinion dissemination algorithm, constrained by dynamic equations, which enables a more realistic macro-level simulation results.
    \item We combine opinion dynamics with the concept of SIR, effectively simulating the phenomenon of the public’s attitude gradually changing toward neutrality after the occurrence of news.
    \item We performed extensive experiments on accurate real-world Weibo data using ChatGLM, showcasing the superior accuracy of our approach when compared to traditional ABM and LLM-based algorithms.
\end{itemize}

\section*{Method}
\subsection*{Offline News}
For news to begin spreading on social network, it must use real-world events that have genuinely taken place. These events are referred to as ``offline news''.

In our study, we specifically selected offline news characterized by a reversal of outcomes between their initial and concluding phases, hereafter referred to as ``reversal news''. Such news typically begins with influential opinion leaders posting unverified claims (rumors), leading to a one-sided public opinion. After a period of time, official authorities release the truth or investigation results, causing a sudden change in public opinion to the opposite. Over time, as the public gradually regains rational judgment, the average attitud tend toward neutrality.

When using LLMs to simulate opinion leaders, we only adopt three attitude states: 1, 0, and -1\cite{chuang-etal-2024-simulating}. This allows us to effectively simulate the process: \textbf{supporting A $\rightarrow$ (potential attitude decay) $\rightarrow$ news reversal $\rightarrow$ supporting B $\rightarrow$ attitude decay}. For LLMs, generating clear opinions (agree/disagree) is easier than specific values (0.7), so -1, 0, 1 are more suitable. Additionally, we incorporate the CA model to generate the final opinion leaders attitude.

\subsection*{Data Analysis}
\subsubsection*{Dataset}
We utilized event keywords on the Chinese social media platform Weibo as crawling markers, collecting related posts and comments daily for each event. In total, we gathered 255176 posts. Our data collection focuses on four highly ``reversal news'': the ``Pangmao'' suicide event (133834), the secondary school girl ``Jiangping'' ranking 12th in a global math competition (98471), the ``Qingdao'' subway assault (6331), and the ``Dianduji'' girl stockpile cancer video incident (16540). We designed a few-shot LLM-based agent to evaluate the attitudes of all of the statements on a scale of [-1, 1]. Attitude values are recorded daily, with the total duration spanning from the day the event occurs. Moreover, the news began to circulate until the overall real attitudes stabilized and approached neutrality. The attitude and action models are based on GLM4 with the same prompt to maintain logical consistency.

The four datasets exhibit a progressively decreasing level of internet discussion intensity, representing events with four different levels of discussion popularity. These data allows for a more comprehensive evaluation of the effectiveness of our simulation method. Detailed data on opinion leaders and opinion followers can be found in Table \ref{tab:dataset_info}.

The following are time segments extracted during the experiment.

\begin{figure}[htbp]
    \centering

    \subcaptionbox{Pangmao Incident\label{pangmaoheat}}{%
        \includegraphics[width=0.45\textwidth]{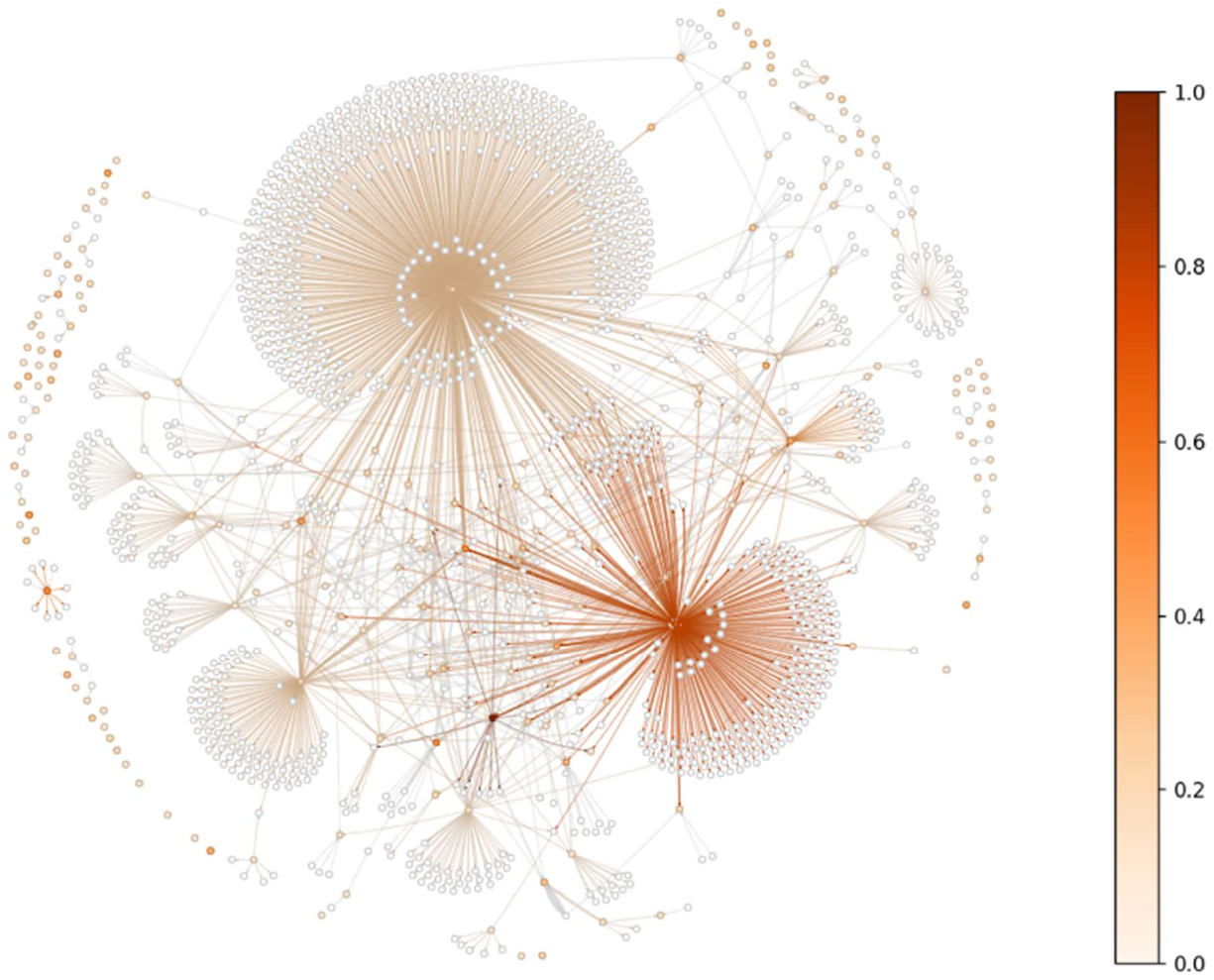}
}
    \hspace{0.04\textwidth} % fine-tuning the space between images
    \subcaptionbox{Jiangping Incident\label{jiangpingheat}}{%
        \includegraphics[width=0.45\textwidth]{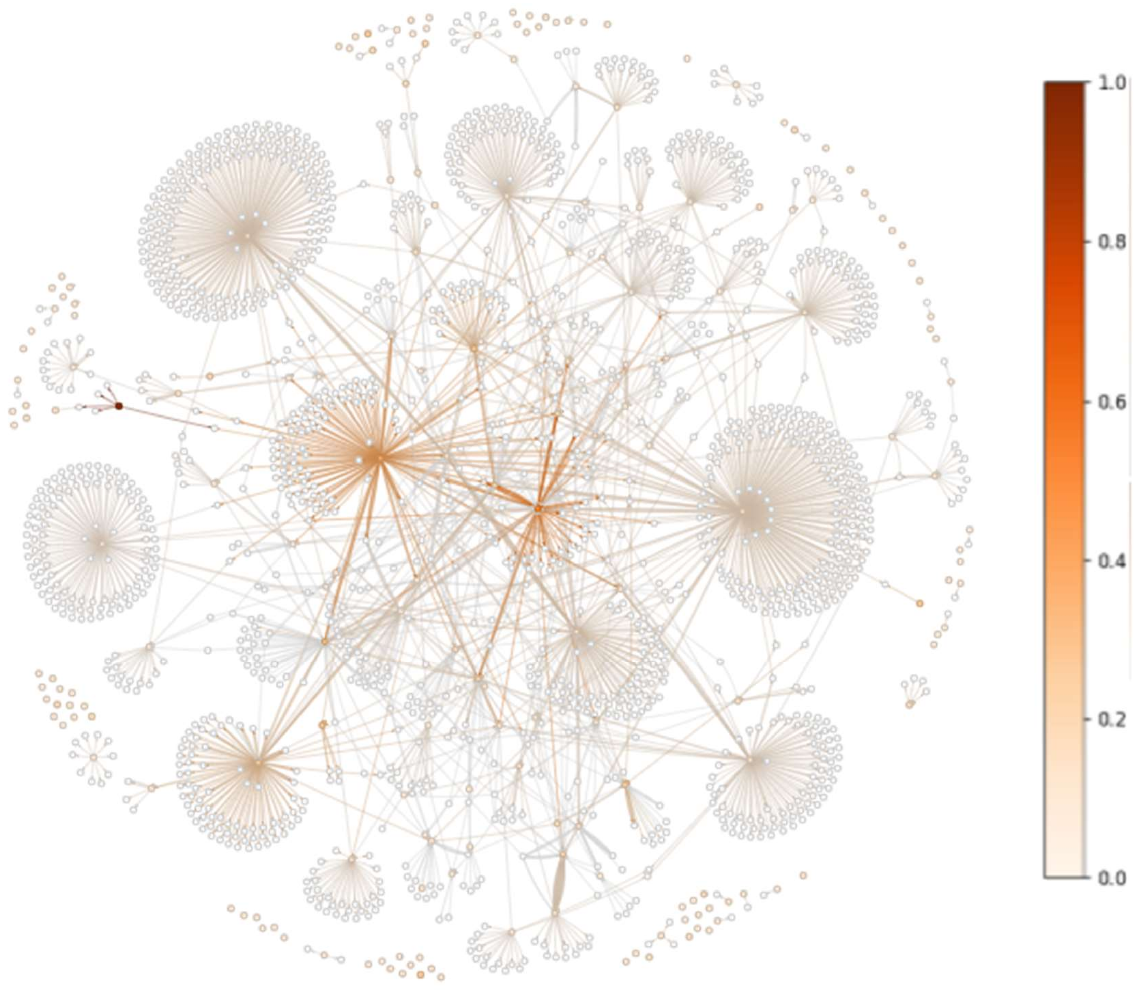}}
    \\
    \subcaptionbox{Qingdao Incident\label{qingdaoheat}}{%
        \includegraphics[width=0.45\textwidth]{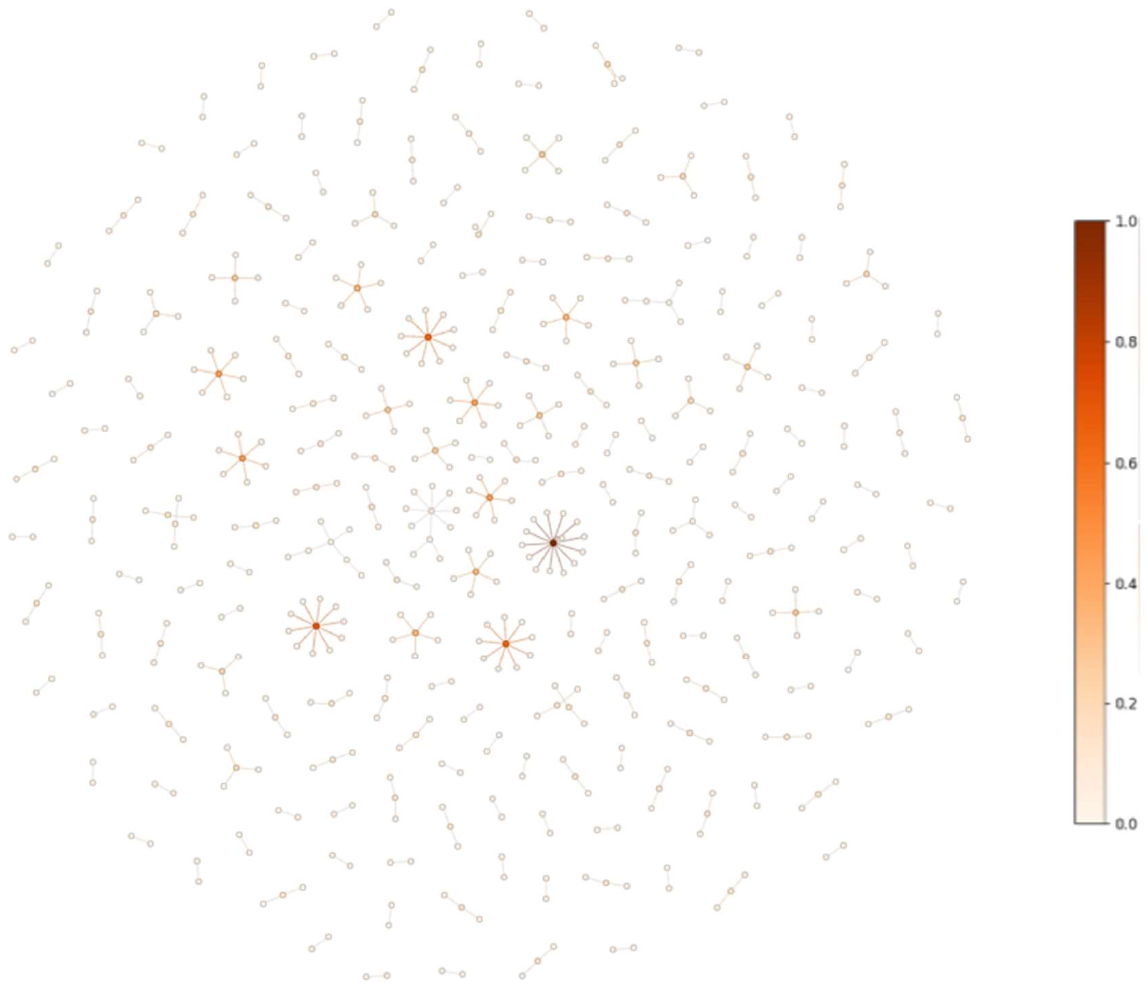}}
    \hspace{0.04\textwidth} % fine-tuning the space between images
    \subcaptionbox{Dianduji Insident\label{diandujiheat}}{%
        \includegraphics[width=0.45\textwidth]{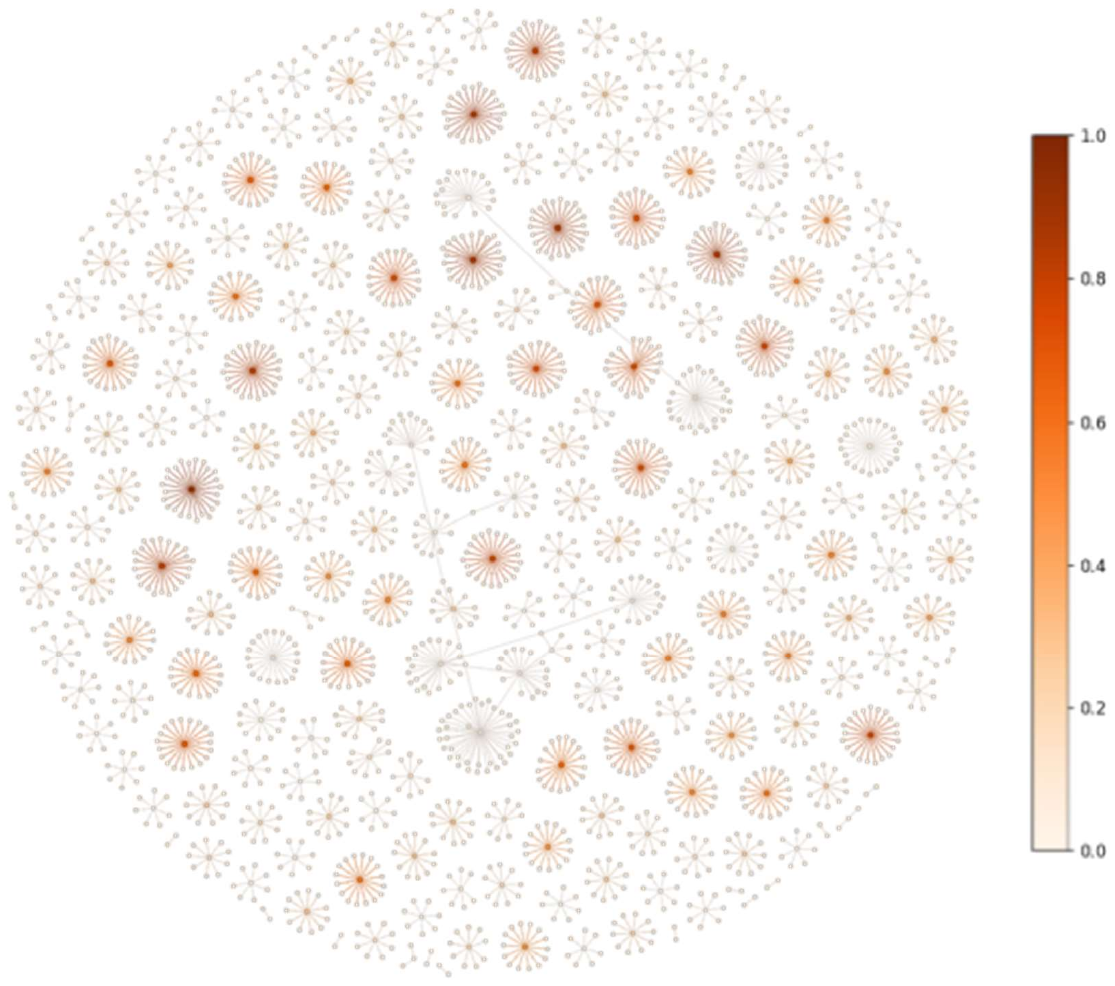}}

      \caption{Event Relationship Network Diagram. The core of the network represents opinion leaders, while the branches represent opinion followers. The deeper the color, the stronger the influence.}
           \label{Event Relationship Network Diagram}
\end{figure}

\textbf{Pangmao Incident} (Figure \ref{pangmaoheat}): On April 11, 2024, a 21-year-old game coach, ``Pangmao'', tragically died after jumping off the Chongqing Yangtze River Bridge. On May 2, 2024, Pangmao's sister shared screenshots of their chat history with his girlfriend, sparking widespread debate. On May 19, 2024, the Chongqing Public Security Bureau's Nanan District branch issued an official report on the April 11 incident through the ``Safe Nanan'' platform, revealing that Pangmao’s sister had intentionally guided negative public opinion, which posed a threat to the social network environment.

\textbf{Jiangping Incident} (Figure \ref{jiangpingheat}): On June 13, 2024, the final list of the 2024 Alibaba Global Mathematics Competition was announced. A 17-year-old vocational student, Jiang Ping, from Lianshui Vocational School in Jiangsu Province, placed 12th, which quickly attracted widespread attention. On June 21, 2024, Zhao Bin, a master student from Peking University, publicly accused Jiang Ping of cheating.

\textbf{Qingdao Incident} (Figure \ref{qingdaoheat}): On August 7, 2024, in Qingdao, Shandong, a young man was allegedly slapped and punched on the subway by an elderly man after refusing to give up his seat, leading to bleeding from his mouth and nose, which sparked heated discussion. On August 8, 2024, Heilongjiang News reported that the young man involved appeared and explained the incident, stating in a video, \textit{``The old man complained because I didn’t offer him my seat, but I only give seats to women, why should I give it to him''} On August 12, 2024, the police issued a statement revealing that the young man had first insulted the elderly man, and the incident had nothing to do with ``giving up the seat''.

\textbf{Dianduji Incident} (Figure \ref{diandujiheat}): On the evening of February 29, 2024, ``Point Reading Machine Girl'' Gao Junyu posted a video on social media stating that she had been diagnosed with a brain tumor and shaved her head. In the following 11 days, her account posted eight videos documenting her surgery and recovery process. On March 11, 2024, some netizens noticed that the clothing in previous videos was inconsistent with the season, suspecting they were ``pre-recorded videos.'' On March 12, 2024, the Internet Illegal and Harmful Information Reporting Center of Yuhang District, Hangzhou, issued a statement, confirming that the videos were filmed in September 2023 and edited starting in February 2024.

Based on the results in Figure \ref{Event Relationship Network Diagram}, the ratio of opinion leaders to followers can be further obtained, as shown in Table \ref{tab:dataset_info}. 
\begin{table}[htbp]
\centering
\begin{tabular}{lccc}
\toprule
Dataset & Opinion Leaders & Opinion Followers & Ratio \\
\midrule
Pangmao  & 1000  & 8890  & 1:9   \\
Jiangping & 200  & 1872  & 1:9   \\
Qingdao  & 20   & 186   & 1:9   \\
Dianduji & 40   & 386   & 1:9   \\
\bottomrule
\end{tabular}
\caption{The ratio distribution of Opinion Leaders and Opinion Followers across different datasets.}
\label{tab:dataset_info}
\end{table}

\subsubsection*{Analysis}\label{analysis}
\begin{figure}[htbp]
    \centering

    \subcaptionbox{Pangmao Incident\label{pangmao}}{%
        \includegraphics[width=0.40\textwidth]{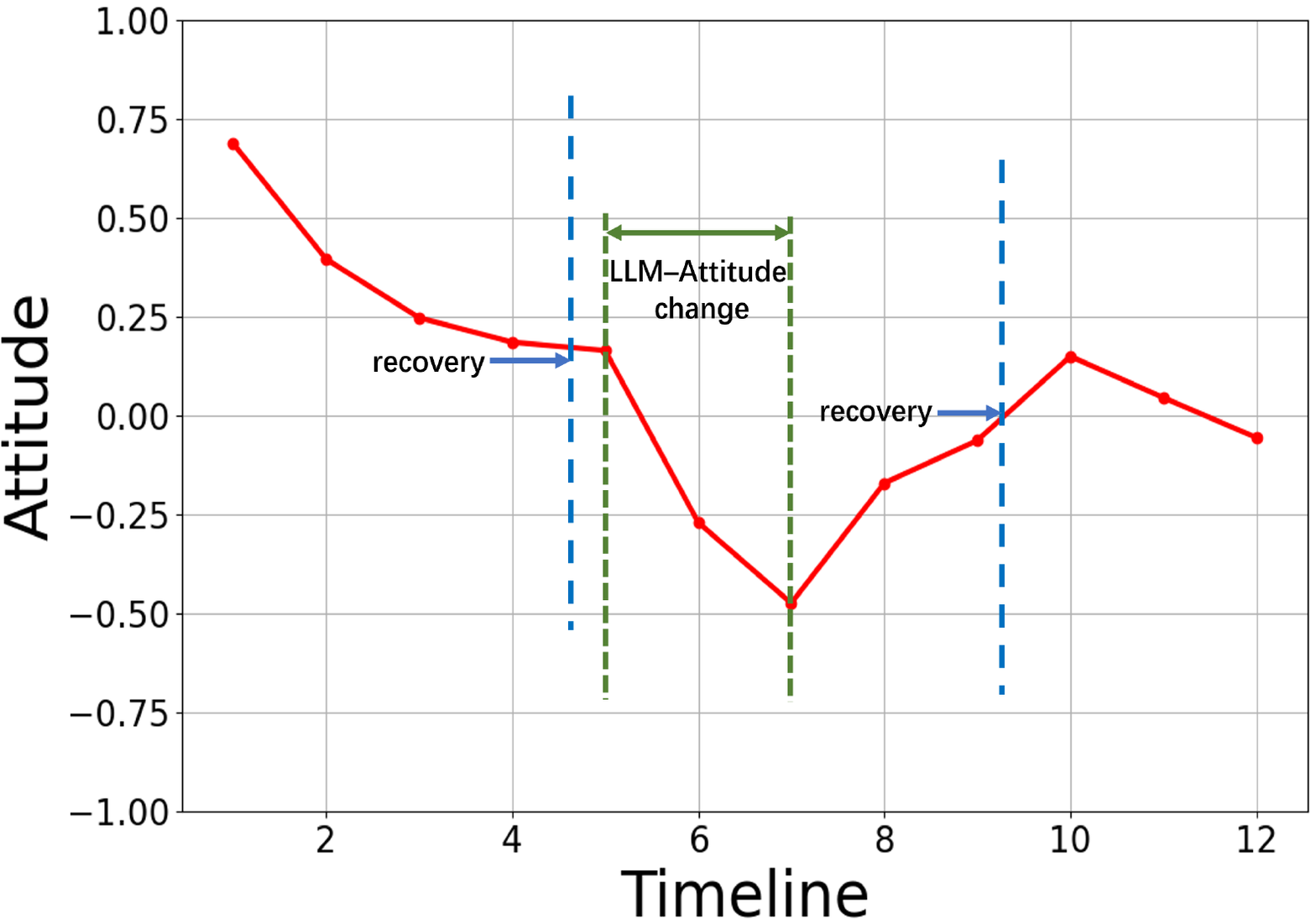}
}
    % \hspace{0.04\textwidth} % fine-tuning the space between images
    \subcaptionbox{Jiangping Incident\label{jiangping}}{%
        \includegraphics[width=0.4\textwidth]{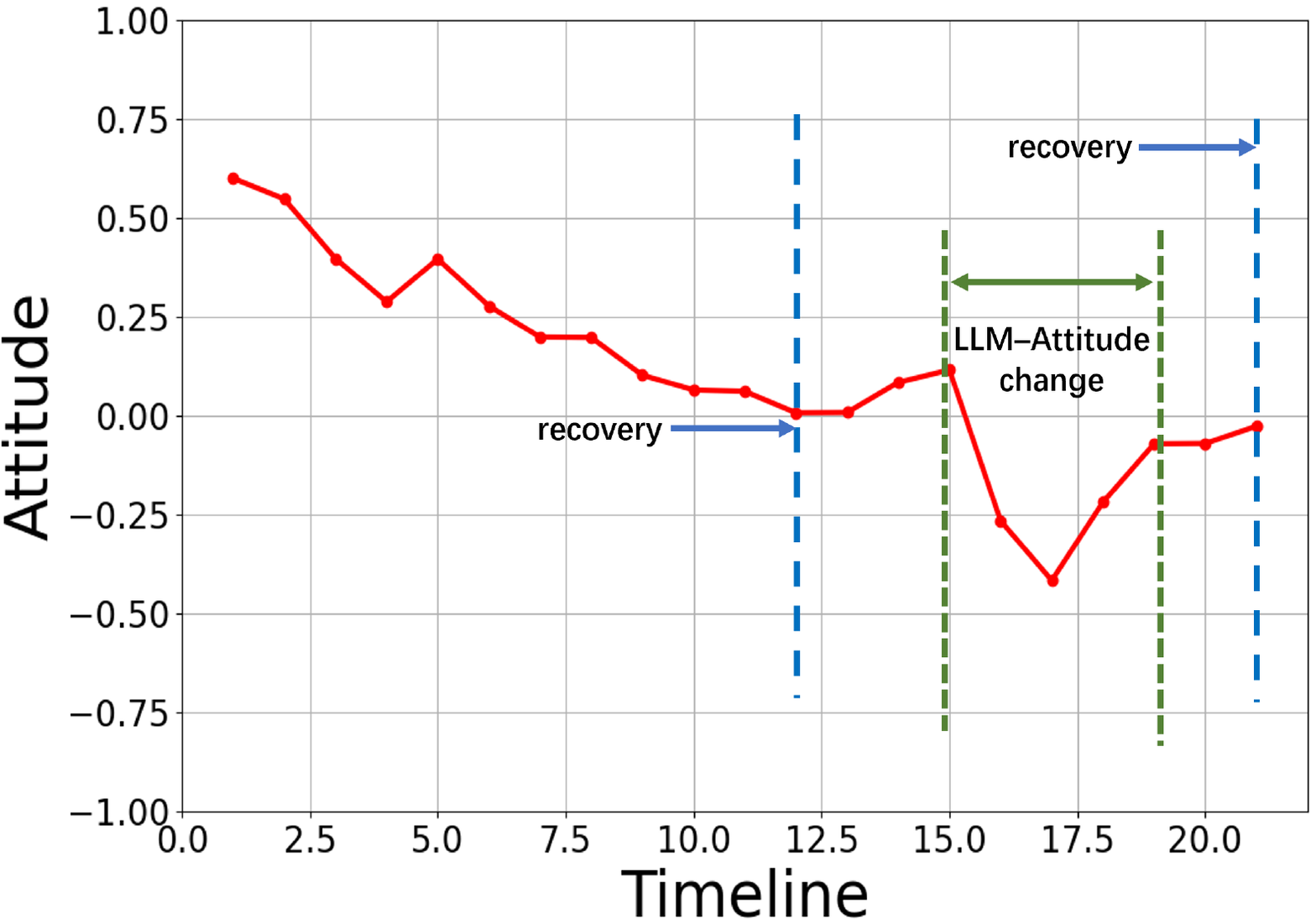}}
    \\
    \subcaptionbox{Qingdao Incident\label{qingdao}}{%
        \includegraphics[width=0.4\textwidth]{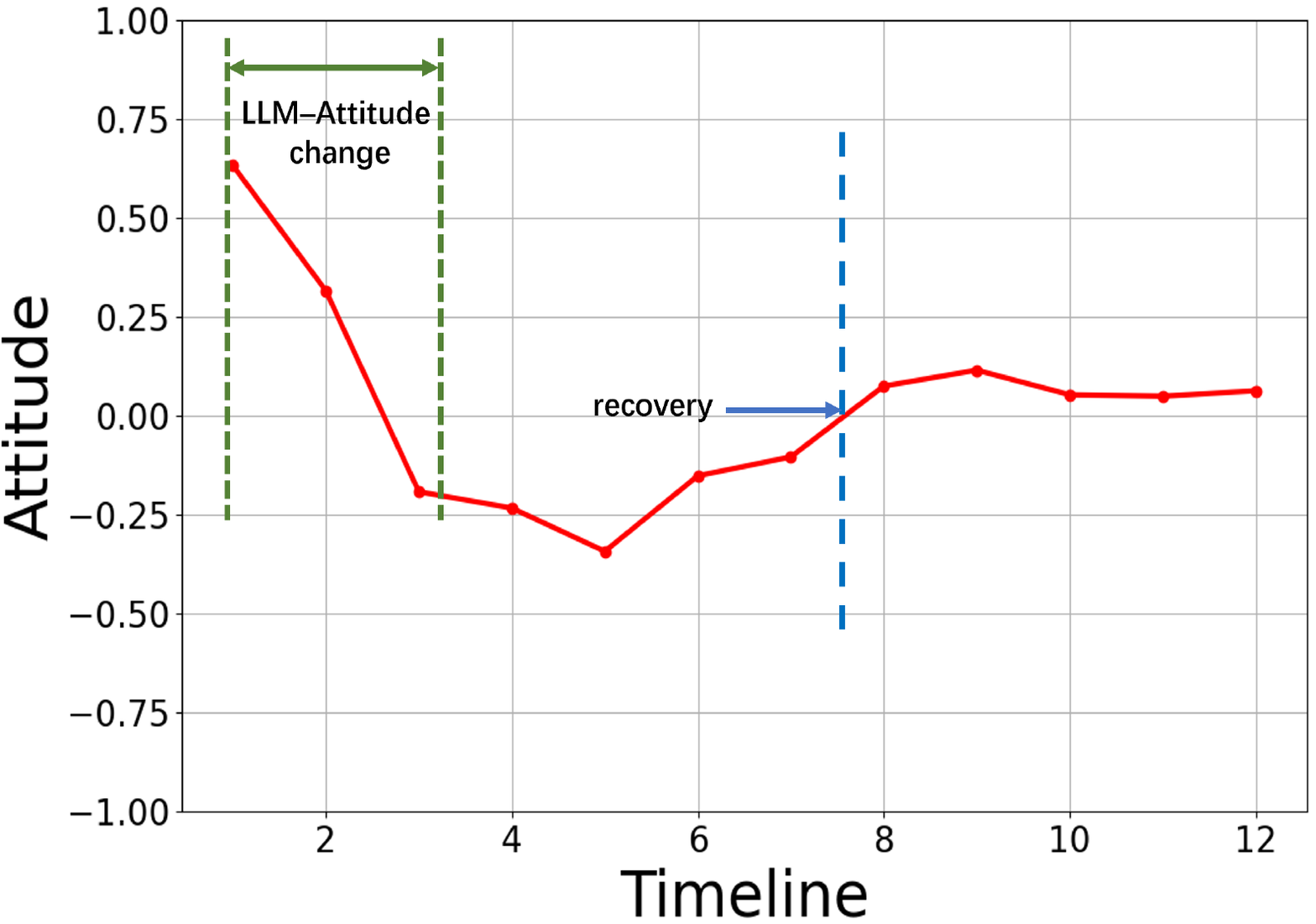}}
    % \hspace{0.04\textwidth} % fine-tuning the space between images
    \subcaptionbox{Dianduji Insident\label{dianduji}}{%
        \includegraphics[width=0.4\textwidth]{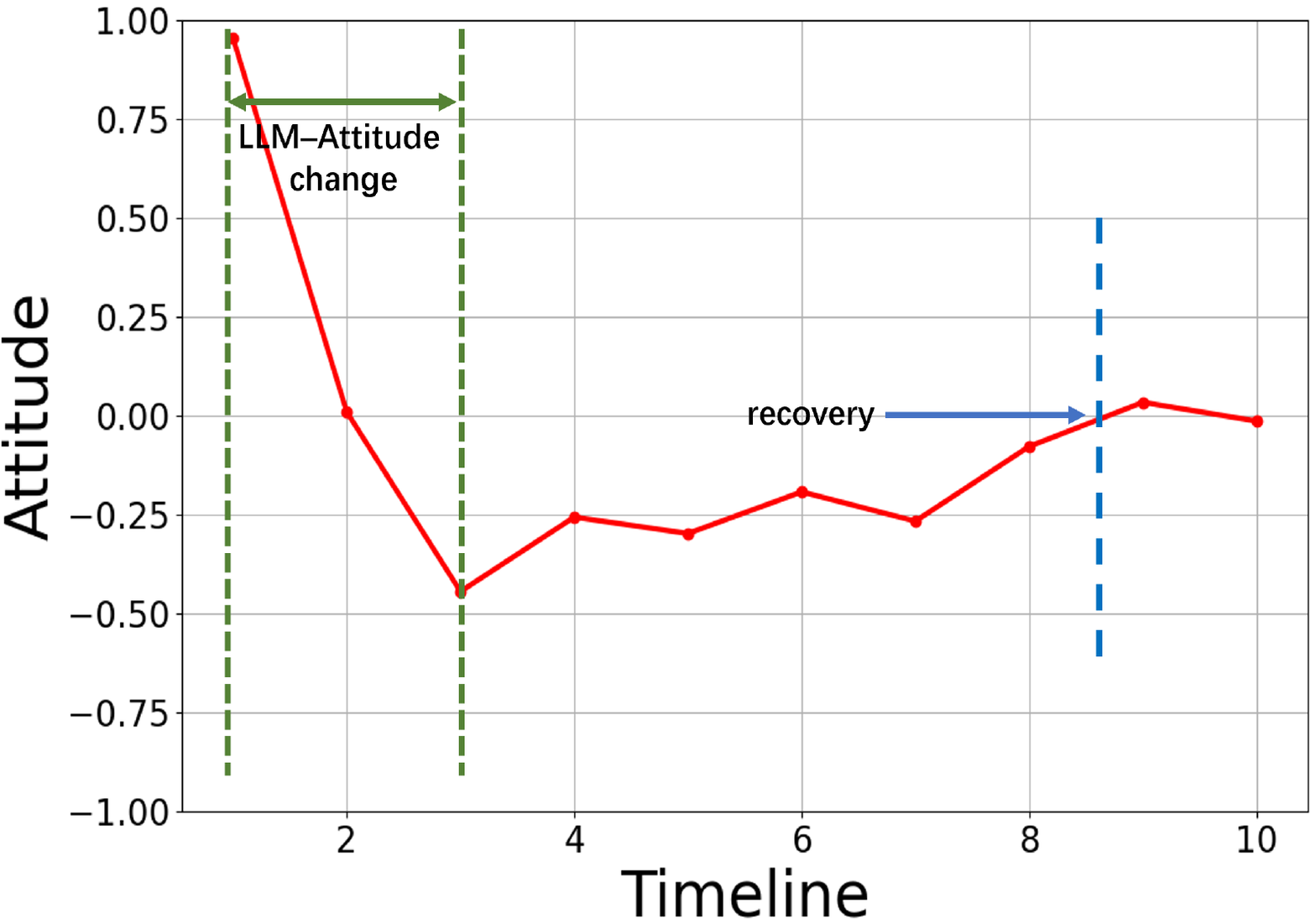}}

      \caption{The red lines represent actual data (a) Pangmao Incident (b) Jiangping Incident (c) Qingdao Incident (d) Dianduji Incident. The labeled parts in the figures are the key areas that we need to simulate using the models.}
           \label{fig:true curve}
\end{figure}

By analyzing the opinion changes with time, they exhibit the following two characteristics:

1. \textbf{Large variance:} The contrast between rumors and the truth leads to a significant change in the attitude. In the Dianduji incident (Figure \ref{dianduji}), the extreme variance of the attitude reached 1.4, indicating that more than half of the people underwent a complete reversal.

2. \textbf{Self-decay:} After the news spreads, we noticed that opinion rises to a certain threshold, then stops growing, gradually declines, and hovers around zero. This trend is often captured in real news through neutral sentiments like  ``questioning authenticity'' or ``both sides might have issues''. For example, in the Jiangping incident (Figure \ref{jiangping}) timeline from day 1 to day 15, support dropped from 0.6 on the first day to 0.1.

To better simulate the aforementioned characteristics, compared to traditional ABM, LLMs can effectively handle large variance, thereby simulating the trend of attitudes in response to significant attitude changes. Additionally, by incorporating the ``recovery'' concept from the SIR model, we can simulate the phenomenon of self-decay.
For more details, please refer to Figure \ref{pangmao}, \ref{jiangping}, \ref{qingdao}, and \ref{dianduji}.

\subsection*{Workflow}
\begin{figure}[htbp]
\centerline{\includegraphics[width=\textwidth]{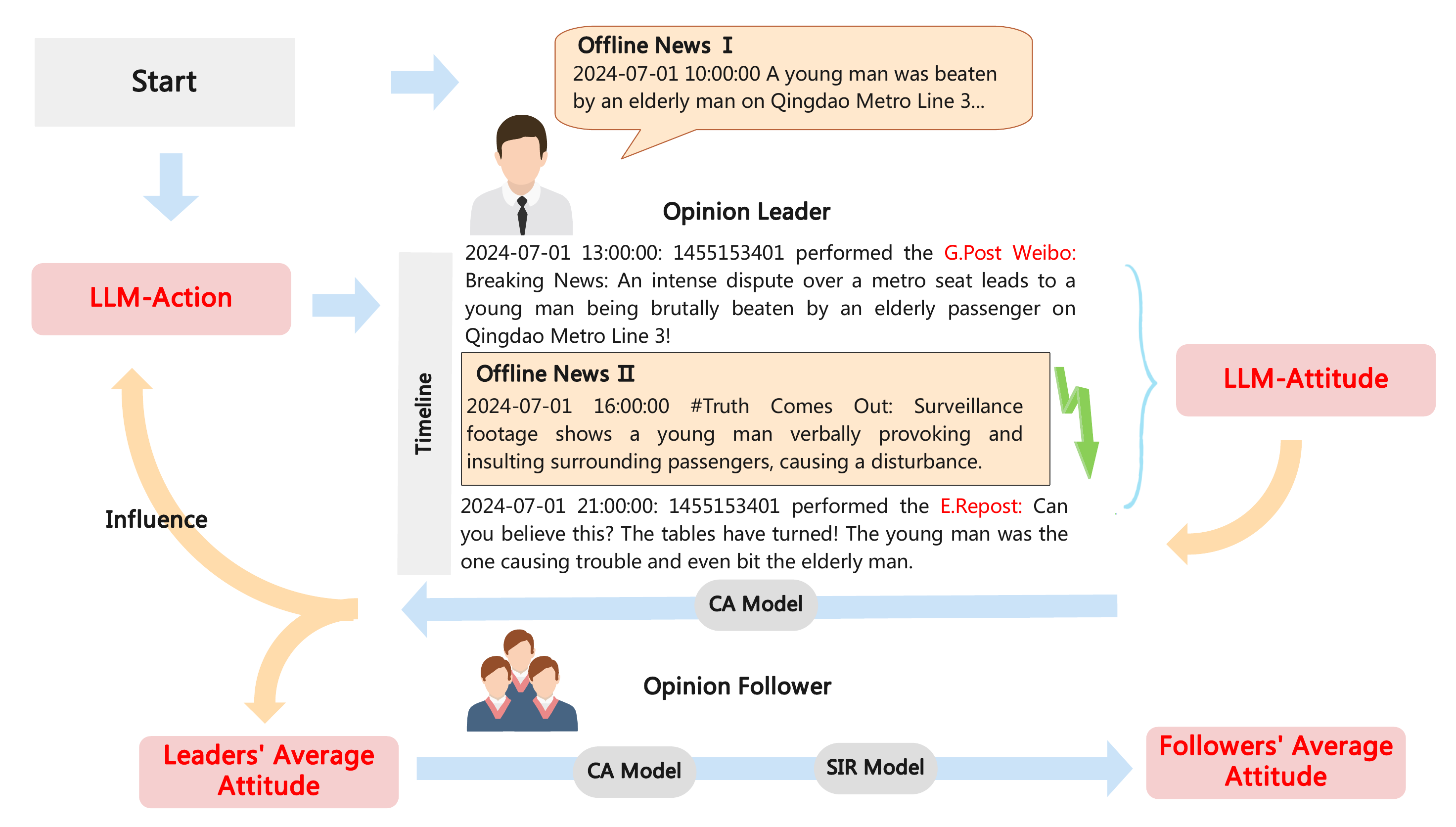}}
    \caption{Workflow of the FDE-LLM Simulation: Cooperation of Opinion Leaders and Opinion Followers, Role-Playing Mechanism, and Integration of CA and SIR Models.}
    \label{fig1 overview above}
\end{figure}

Figure \ref{fig1 overview above} illustrates the overall frameworkof the FDE-LLM algorithm. We developed a simulated social platform by python where opinion leaders possess the ability to take actions (such as commenting and reposting), while opinion followers are influenced by the opinion leaders and followers. In addition to being able to read each other's actions (comments and reposts), opinion leaders will also read offline news at the beginning and end of the simulation, representing the initiation and conclusion (the revelation of the truth) of a ``reversal news''. During the intervals between offline news releases, they will engage in discussions and changes in opinions.

\begin{figure}[htbp]
    \centerline{\includegraphics[width=0.8\textwidth]{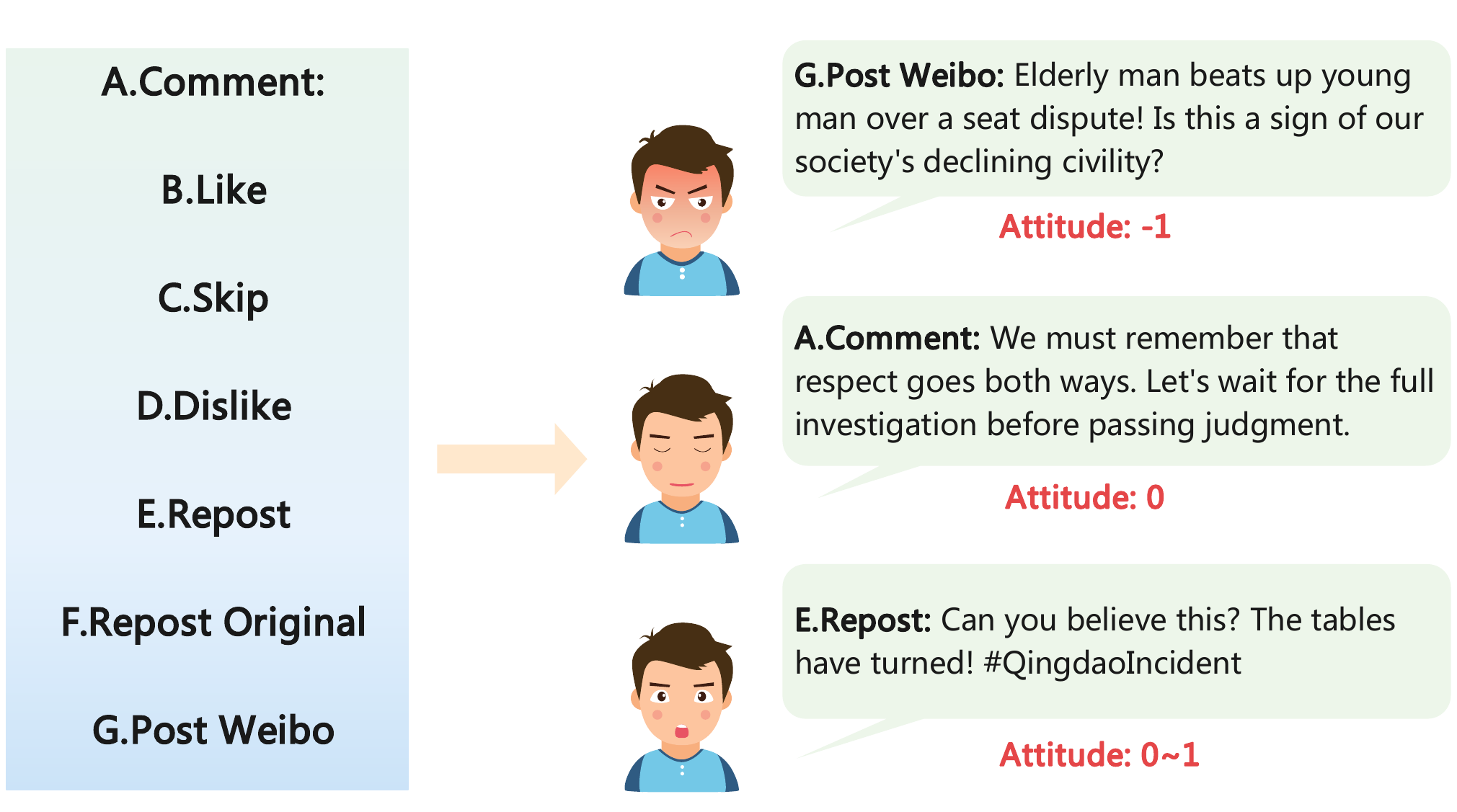}}
    \caption{Actions and Attitudes. The left side displays the types of actions that LLM-Action can select.  In contrast, the right side illustrates the specific behaviors related to each action type and the associated scoring of attitudes by LLM-Attitudes.}
    \label{fig2 action}
\end{figure}

Firstly, the opinion leaders, receive the first offline news, which is inaccurate, misleading, and provocative, and take action (Figure \ref{fig2 action}). LLM-Attitude assesses the subsequent attitudes toward these actions. The results are then separately relayed to opinion leaders (simulated by LLM with CA) and opinion followers (simulated by CA with SIR). Leaders will treat the attitude processed by the CA model as the current attitude value. In the next round, the current attitude value will serve as one of the influencing factors in determining the action choice (Figure \ref{fig3 agent design}). Followers will obtain their own attitude value for the current round by processing Leaders' current attitude value through both the CA and SIR models.

Each round is ``one hour'', and we take the time of truth exposure in reality as the reference, releasing the offline news containing the truth at that point in time. The followers represent the general public, reflecting the direction of public opinion. Therefore, when evaluating attitude, we will use the opinion followers. By comparing the attitudes before and after the truth exposure with the real attitudes, we can assess the effectiveness of the simulation.
\subsection*{Definition}

\subsubsection*{Opinion Leader Agent}

Opinion leaders are initiators of opinion in social networks. Positive opinion leaders typically play a constructive role in key moments, while negative opinion leaders may spread rumors, mislead others, or even disrupt social stability. Therefore, the influence of opinion leaders is dual in nature, as it can either guide the group toward positive goals or lead to negative consequences. 

According to the real-world dataset, we select and model opinion leader based on users' profile (such as critics, celebrities, official media accounts, etc.) (Figure \ref{Event Relationship Network Diagram}, Table \ref{tab:dataset_info}). We maintained an opinion leader to follower ratio of 1:9, ensuring a realistic distribution of influence within the simulation.

These agents consciously spread or diffuse their opinion to followers and guide their perspective; their opinions are not influenced by opinion followers, and there is interaction of opinions only between leaders with similar attitudes. We control this through attitude difference thresholds. As a result, we obtain multiple groups of leaders with distinguishable attitudes.

\subsubsection*{Opinion Follower Agent}

We define an individual who does not possess all of the above characteristics is referred to as an opinion follower. In reality, opinion followers lack a clear target opinion. 

In our agent framework, during the process of updating their perspectives, they are influenced by the \textbf{similar} opinions of opinion leaders as well as other opinion followers.

\subsubsection*{LLM-Action}
The LLM-Action module predicts the actions taken by opinion leaders after reading news. Based on the profiles described by Xinyi Mou et al. \cite{21mou2024unveiling}, we have assigned social personalities to opinion leaders. They are characterized as talkative and provocative, which makes their behavior more radical and reduces the likelihood of neutrality. This setup is beneficial for the subsequent LLM-Attitude module to evaluate the agent's stance.

\begin{figure}[htbp]
\centerline{\includegraphics[width=0.7\textwidth]{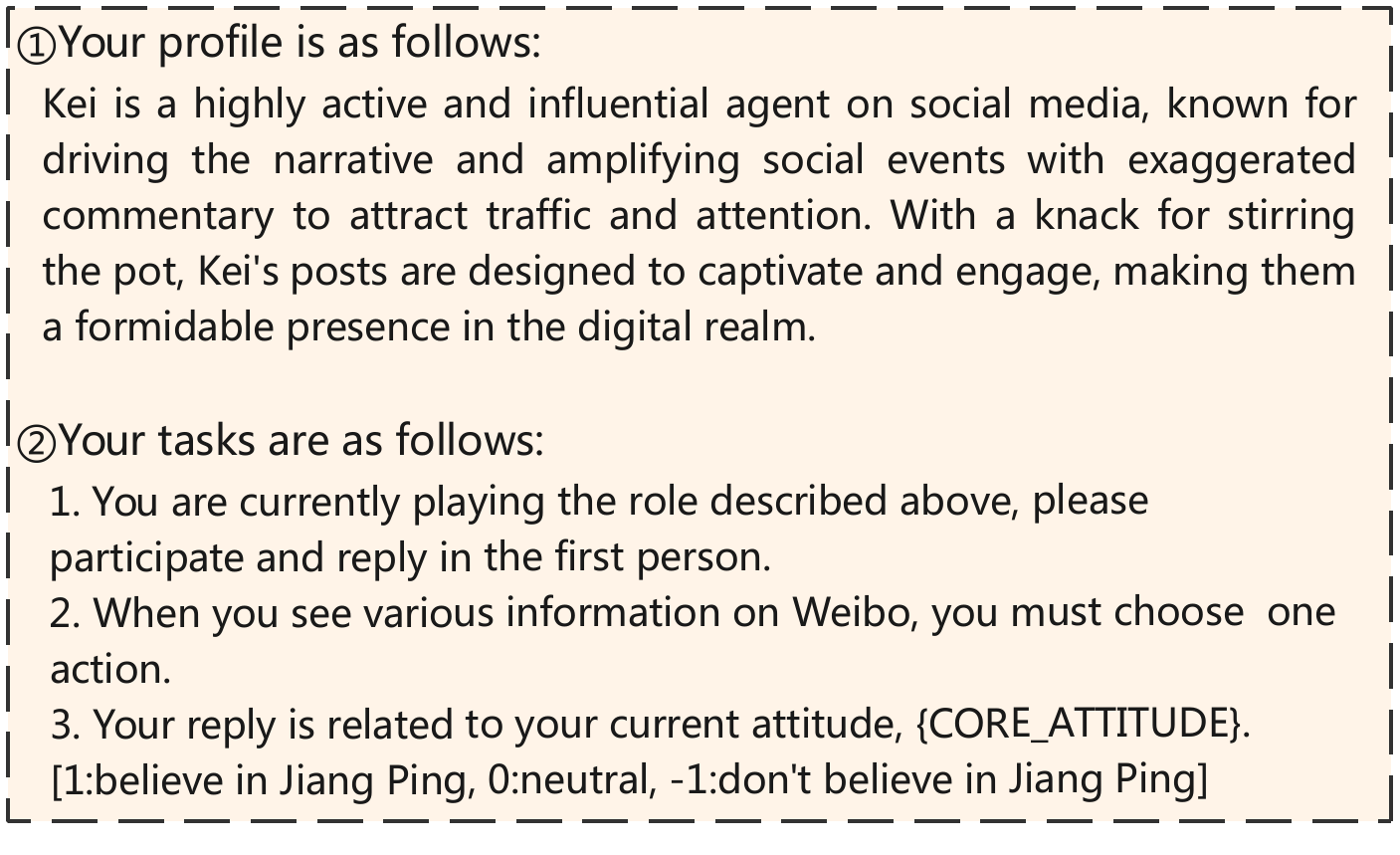}}
    \caption{Agent Profile. This is an example of an active and influential social media agent with narrative amplification and exaggerated commentary to engage the audience. The profile is summarized from the actual dataset, based on the target user's reviews and posts}
    \label{fig3 agent design}
\end{figure}

To ensure the coherence of actions and avoid the erroneous scenario of ``flip-flopping'' between support and opposition, the agent needs to consider the current attitude value when making actions in each round (Figure \ref{fig3 agent design}).

\subsubsection*{LLM-Attitude}
The model also leverages the established framework to build agents by designing targeted prompts for different events and facilitating objective and unbiased attitude evaluations. By employing the same LLM for both action and attitude,  we guarantee that LLM-Attitude and LLM-Action maintain aligned values, which accounts for the uniformity in the generated attitudes.

We designed the agents to select only three values: 1 for support, 0 for neutrality, and -1 for opposition (supporting the opposing side). This design is based on the characteristics of ``reversal news'', which often highlights extreme polarization of opinions. In such cases, the attitudes of groups toward rumors and truths tend to be diametrically opposed before eventually stabilizing. Therefore, using only 1, 0, and -1 is sufficient to reflect the entire process of attitude evolution. Additionally, considering the uncertainty of LLMs, if there are too many intermediate values, the LLM might produce significantly different answers for similar variations of the same action, leading to incorrect attitude judgments.

\subsection*{Models}
\subsubsection*{Opinion Leaders: LLM with CA}

LLMs are effective in modeling scenarios where significant changes in attitude occur, as they can handle large variation in opinion. However, they tend to converge rapidly to a stable value, lacking the ability to simulate ongoing changes and interactions. In contrast, CA cannot easily capture sudden changes in events but excel in simulating the effects of mutual influence between individuals. 

By combining these two models, we can leverage the strengths of each: LLMs for handling large-scale attitude changes, and CA for modeling the intricate interactions and continuous dynamics within a group. To capture the micro-level interactions between opinion leaders, we map their opinion states onto a CA grid\cite{Hu2018CellularA}, where each cell represents an agent. We chose a grid-based model over a fully connected network to balance realism and practicality. While fully connected networks capture complex, long-range interactions, they require extensive data (e.g., user follow lists) and computational resources. The grid model assumes local interactions, aligning with event-driven dynamics where most of opinion leaders often exert localized influence within communities. It fosters local consensus while preserving global diversity, suitable for scenarios without full consensus \cite{suo2008dynamics, li2013using, 6108515}. Additionally, it offers data efficiency, privacy protection, and scalability by avoiding sensitive data and enabling flexible group size adjustments. The CA model is shown in Equation \ref{ca1}, \ref{ca2}.
\begin{equation}
\label{ca1}
    O_i^{t+1}=r\cdot O_i^t+w\cdot\sum_{j\in X_i}\ T_{ij}^t
\end{equation}
\begin{equation}
\label{ca2}
    T_{ij}^t = \begin{cases} 
      O_j, & r = 0 \\
      0, & r = 1 \text{ or } \left| O_j^t - O_i^t \right| > \epsilon \\
      \left( O_j^t - O_i^t \right) \cdot \sqrt{r \cdot |O_j^t|}, & r \neq 1, 0 \text{ and } \left| O_j^t - O_i^t \right| \leq \epsilon
   \end{cases}
\end{equation}
where $O_i^{t+1}$ represents the updated opinion of the individual opinion leader $i$ at time $t+1$. $O_i^t$ is the opinion of the individual opinion leader $i$ at time $t$; $r$ is the retention factor that indicates how much of the current opinion is retained; $w$ refers to the influence coefficient that determines the impact of neighboring opinions. At the same time,  $\sum_{j\in X_i} T_{ij}^t$ represents the summation of influences $T_{ij}^t$ from all neighbors $j$ in the neighborhood $X_i$ at time $t$. This formula models the evolution of an individual opinion leader's opinion over time, accounting for personal retention and the influence of surrounding opinions.

\textbf{Mechanism 1 (Opinion persistence):} The retention factor $r \in [0,1]$ quantifies cognitive consistency through diagonal dominance in the update matrix. 

\textbf{Mechanism 2 (Thresholded influence):} The tolerance threshold $\epsilon$ implements confirmation bias by nullifying influences beyond $|\Delta O| > \epsilon$. 

\textbf{Mechanism 3 (Non-linear influence):} The term $\sqrt{r\cdot|O_j^t|}$ captures the effect of neighboring agents' opinions on an individual's attitude.

We designed the following formula to constrain the attitudes generated by LLM using CA (Equation \ref{llm+ca}):  
\begin{equation}
\label{llm+ca}
\begin{split}
O_i^{t+1} = \mathrm{clip}\Bigg( & \alpha \cdot \left(r \cdot O_i^t + w \cdot \sum_{j \in N_i} T_{ij}^t\right) + (1 - \alpha) \cdot \text{LLM}, -1, 1 \Bigg)
\end{split}
\end{equation}
We outline the $clip(-1, 1)$ function to ensure that the updated opinion values remain within the range of $[-1, 1]$. This framework prevents the opinion values from exceeding reasonable limits, ensuring the stability and rationality of the model results. In addition, an $\alpha$ fusion coefficient is introduced to determine the relative influence of the LLM and the CA model.  The pseudocode is shown in Algorithm \ref{ps:llm+ca}.

\begin{algorithm}[htbp]
\caption{LLM + CA Model\label{ps:llm+ca}}
\begin{algorithmic}[1]
\STATE \textbf{Initialize Parameters:}
\STATE Initialize the opinion values $O_i^0$ for $i = 1, 2, ..., N$
\STATE Set parameters: retention factor $r$, influence coefficient $w$, threshold $\epsilon$, and fusion coefficient $\alpha$

\STATE \textbf{For each time step $t$:}
\FOR{each agent $i$}
    \STATE Update opinion using Cellular Automata:
    \[
    O_i^{t+1} = r \cdot O_i^t + w \cdot \sum_{j \in X_i} T_{ij}^t
    \]
    \STATE \textbf{Compute Influence $T_{ij}^t$ as:}
    \[
    T_{ij}^t = 
    \begin{cases} 
      O_j, & r = 0 \\
      0, & r = 1 \text{ or } |O_j^t - O_i^t| > \epsilon \\
      (O_j^t - O_i^t) \cdot \sqrt{r \cdot |O_j^t|}, & r \neq 1, 0 \text{ and } |O_j^t - O_i^t| \leq \epsilon
    \end{cases}
    \]
\ENDFOR

\STATE \textbf{Update Opinion with LLM and CA Fusion:}
\FOR{each agent $i$}
    \STATE Update opinion with LLM and CA fusion:
    \[
    O_i^{t+1} = \text{clip}\left( \alpha \cdot \left( r \cdot O_i^t + w \cdot \sum_{j \in N_i} T_{ij}^t \right) + (1 - \alpha) \cdot \text{LLM}, -1, 1 \right)
    \]
\ENDFOR

\STATE \textbf{Repeat until convergence or stopping criteria are met.}
\end{algorithmic}
\end{algorithm}

\subsubsection*{Opinion Follower: CA with SIR}\label{opinion follower}
The CA model demonstrates strong efficiency in simulating attitude propagation within a group. However, our examination of real datasets revealed an intriguing phenomenon:  self-decay.

To address this challenge, we have integrated part of the SIR model. By introducing a probabilistic decay rule inspired by SIR's recovery concept, we effectively overcome the limitation of unconstrained attitude propagation in the CA model. This adaptation allows opinion followers to regain balance when exposed to aligning with the opinion leader's position, thereby more accurately mirroring the natural decay of attitudes in social news discussions.
The CA-SIR model is as follows:
%\begin{equation}
   $ O_i^{t+1} = O_i^{t+1} \cdot e^{-\lambda \cdot |O_i^{t+1}|} \quad \text{if} \quad \text{random} < \gamma$
%\end{equation}

The fusion model of CA and SIR is described by the following Equation \ref{equ:sirca}.
\begin{figure*}[htbp]
\begin{equation}\label{equ:sirca}
    O_i^{t+1} = \mathrm{clip}\left(\left(r \cdot O_i^t + w \cdot \frac{\sum_{j \in N_i} \left( O_j^t - O_i^t \right) \cdot \sqrt{r \cdot |O_j^t|} \cdot I\left( |O_j^t - O_i^t| \leq \epsilon \right)}{|N_i|}\right) \cdot e^{-\lambda \cdot |O_i^{t+1}|} \cdot I(\text{random} < \gamma), -1, 1\right)
\end{equation}
\end{figure*}

where $|N_i|$ is the number of neighbors around individual $i$: $|O_j^t - O_i^t|$ signifies the opinion difference between neighbor $j$ and individual $i$; $I\left(|O_j^t - O_i^t| \leq \epsilon\right)$ is an indicator function that equals 1 if the opinion difference is within the threshold $\epsilon$, and 0 otherwise. $\lambda$ is the decay rate indicating the natural decay of opinions. $\gamma$ is the recovery rate, representing the probability of opinion recovering.
The pseudocode is shown in Algorithm \ref{ps:ca+sir}.

\begin{algorithm}[htbp]
\caption{CA + SIR Model\label{ps:ca+sir}}
\begin{algorithmic}[1]
\STATE \textbf{Initialize Parameters:}
\STATE Initialize the opinion values $O_i^0$ for $i = 1, 2, ..., N$
\STATE Set parameters: retention factor $r$, influence coefficient $w$, threshold $\epsilon$, decay rate $\lambda$, and recovery probability threshold $\gamma$

\STATE \textbf{For each time step $t$:}
\FOR{each agent $i$}
    \STATE \textbf{Update Opinion using Cellular Automata:}
    \[
    O_i^{t+1} = r \cdot O_i^t + w \cdot \frac{\sum_{j \in N_i} \left( O_j^t - O_i^t \right) \cdot \sqrt{r \cdot |O_j^t|} \cdot I\left( |O_j^t - O_i^t| \leq \epsilon \right)}{|N_i|}
    \]

    \STATE \textbf{Apply SIR Decay (Recovery $\gamma$):}
    \[
    O_i^{t+1} = O_i^{t+1} \cdot e^{-\lambda \cdot |O_i^{t+1}|} \quad \text{if} \quad \text{random} < \gamma
    \]

    \STATE \textbf{Clip the Opinion Value:}
    \[
    O_i^{t+1} = \text{clip}(O_i^{t+1}, -1, 1)
    \]

\ENDFOR

\STATE \textbf{Repeat until convergence or stopping criteria are met.}
\end{algorithmic}
\end{algorithm}

\section*{Experiments}

\subsection*{Experimental Setup}
The GLM4 model was run in a Python 3.10.12 environment utilizing an Intel(R) Xeon(R) Platinum 8160 CPU @ 2.10GHz, a T4 GPU. Our approach focuses on integrating ABM with LLMs to simulate human behavior in opinion dynamics, emphasizing LLMs' role-playing capabilities over advanced reasoning. GLM4 was chosen for its robust performance in generating contextually relevant responses, which is sufficient for modeling opinion leader interactions.

1. \textbf{Parameter settings:} Through experimentation, we found that we could use unified parameters for all of the models on similar reversal news. The parameters were derived through grid search using the ABM algorithm to maximize the correlation coefficient. The specific settings are as follows: For the CA model: opinion resilience \(r = 0.99\), neighborhood influence coefficient \(w = 0.3\), opinion interaction threshold \(\epsilon = 0.5\); for the SIR model: infection rate \(\beta = 0.3\), recovery rate \(\gamma = 0.9\), decay rate \(\lambda = 0.5\). Based on the results in Figure \ref{Event Relationship Network Diagram} and Table \ref{tab:dataset_info}, we set the ratio between opinion leaders and opinion followers as 1:9.

2. \textbf{Model Configurations:} 
(1) LLM: Simulations employ only LLM, without any intervention.
(2) ABM(CA)\cite{Hu2018CellularA}: We use ABM's commonly employed CA model.
(3) ABM(HK)\cite{7rainer2002opinion}: The model employs the ABM's commonly employed HK model.
(4) ABM(PM)\cite{11duggins2014psychologically}: We employ Duggins' Psychologically-Motivated ABM model.
(5) ABM(RLE)\cite{10carpentras2022deriving}: We employ Carpentras' ABM model which based on real life experimen.
(6) LLM+ABM(CA): We simulate opinion leaders through LLM, while ABM uses the CA model to simulate group attitudes.
\subsection*{Indicator}
We evaluate the effectiveness of the models in simulation using two indicators.

1. \textbf{Pearson Correlation Coefficient (Corr.)} is used to measure the linear correlation between the simulated attitude sequence $S=\left(s_1,s_2,\ldots,s_n\right)$ and the real attitude sequence $T=\left(t_1,t_2,\ldots,t_n\right)$. The formula is as follows:
\begin{equation}
r = \frac{\sum_{i=1}^{n}\left(s_i-\bar{s}\right)\left(t_i-\bar{t}\right)}{\sqrt{\sum_{i=1}^{n}\left(s_i-\bar{s}\right)^2} \sqrt{\sum_{i=1}^{n}\left(t_i-\bar{t}\right)^2}}
\end{equation}
where $\bar{s}$ and $\bar{t}$ are the mean values of the simulated and real attitudes, respectively. The Corr. ranges from [-1, 1], and the closer the value to 1, the stronger the linear correlation between the simulated and real attitudes, indicating that the model better captures changes in real attitudes.

2. \textbf{Dynamic Time Warping (DTW)} is used to measure the similarity between the simulated attitude sequence $S$ and the real attitude sequence $T$, especially when the two sequences differ in length or have temporal changes. DTW aligns the two sequences by calculating the minimum cumulative distance. The formula is as follows:
\begin{equation}
DTW(S, T) = \min{\sqrt{\sum_{k=1}^{K} \left(s_{i_k} - t_{j_k}\right)^2}}
\end{equation}
where $\left(i_k, j_k\right)$ signifies the aligned points in sequences $S$ and $T$, and $K$ is the length of the alignment path. A smaller DTW distance indicates a higher similarity between the time series of the simulated and real attitudes, signifying that the model performs better in simulating complex time series patterns.  We can better evaluate the model's accuracy and robustness in dynamic environments using DTW.
\subsection*{Main Result}

\begin{figure}[htbp]
    \centering
  
    \includegraphics[width=0.8\textwidth]{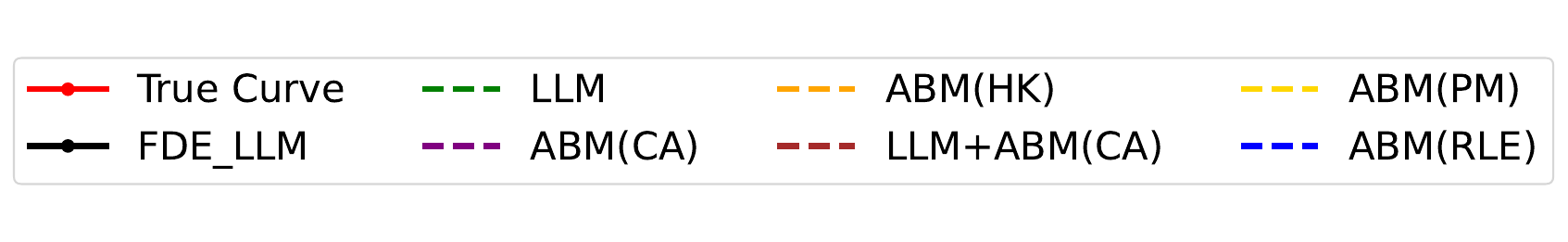}
    \subcaptionbox{Pangmao Incident}{%
        \includegraphics[width=0.45\textwidth]{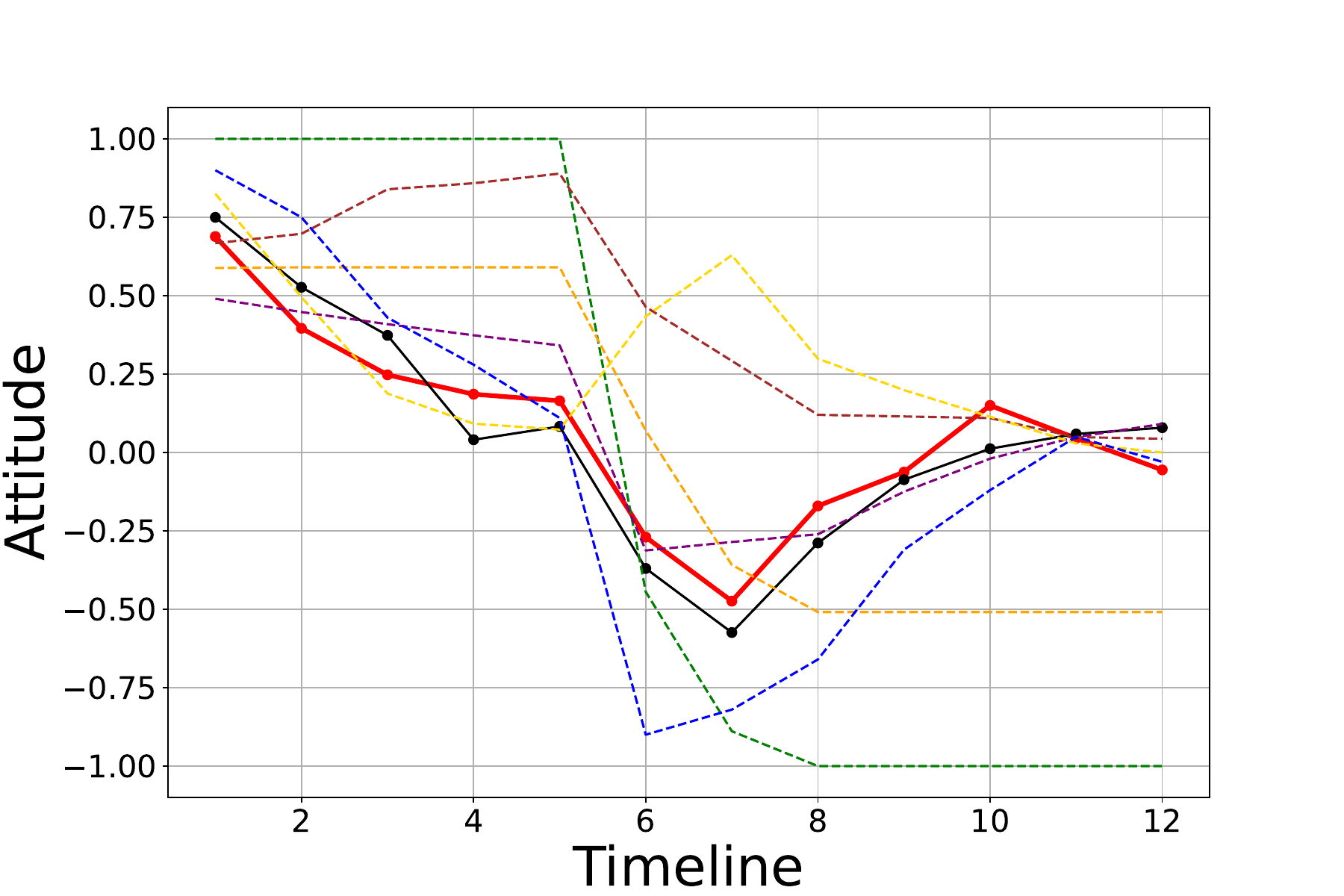}
}
    % \hspace{0.04\textwidth} % fine-tuning the space between images
    \subcaptionbox{Jiangping Incident}{%
        \includegraphics[width=0.45\textwidth]{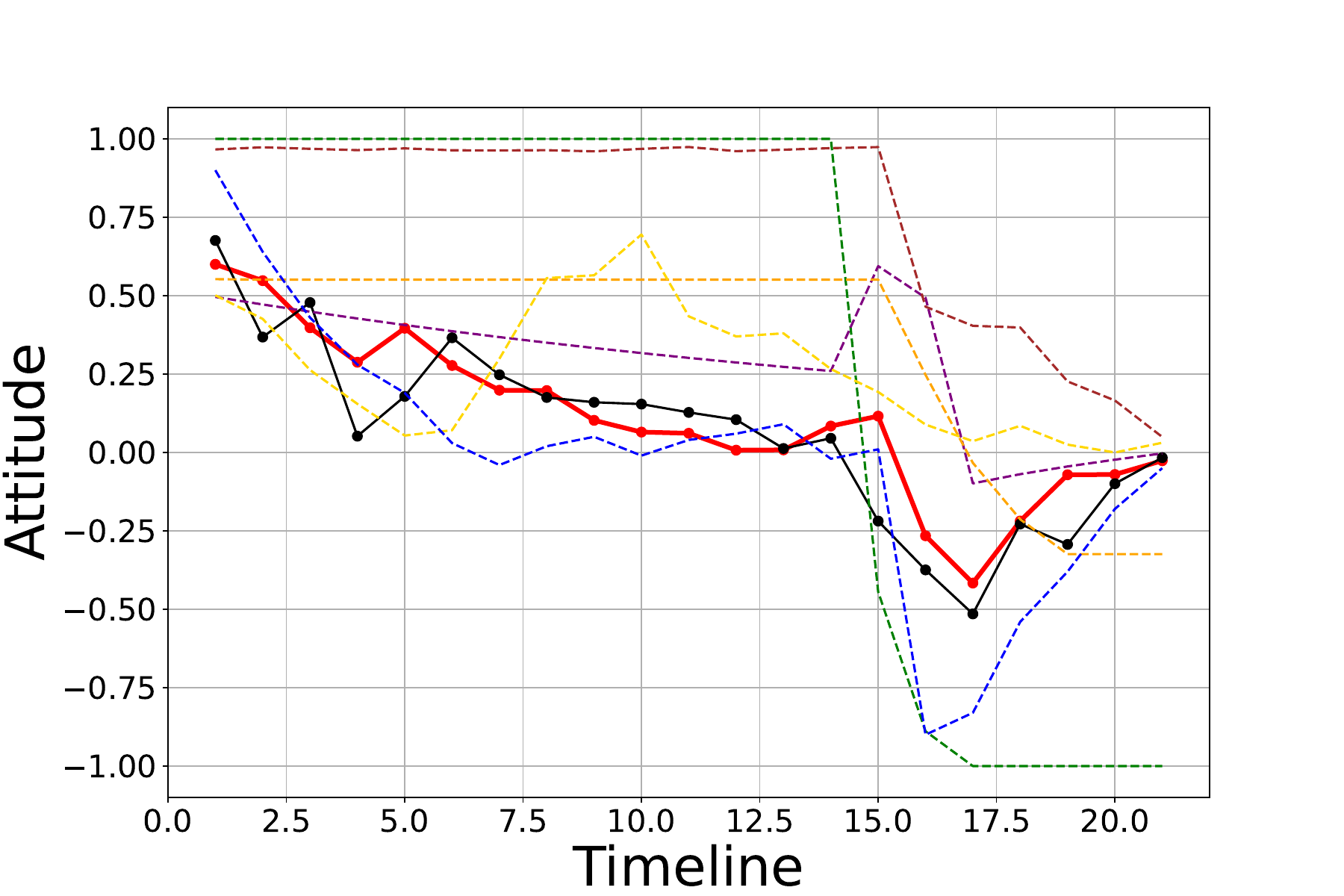}}
    \\
    \subcaptionbox{Qingdao Incident}{%
        \includegraphics[width=0.45\textwidth]{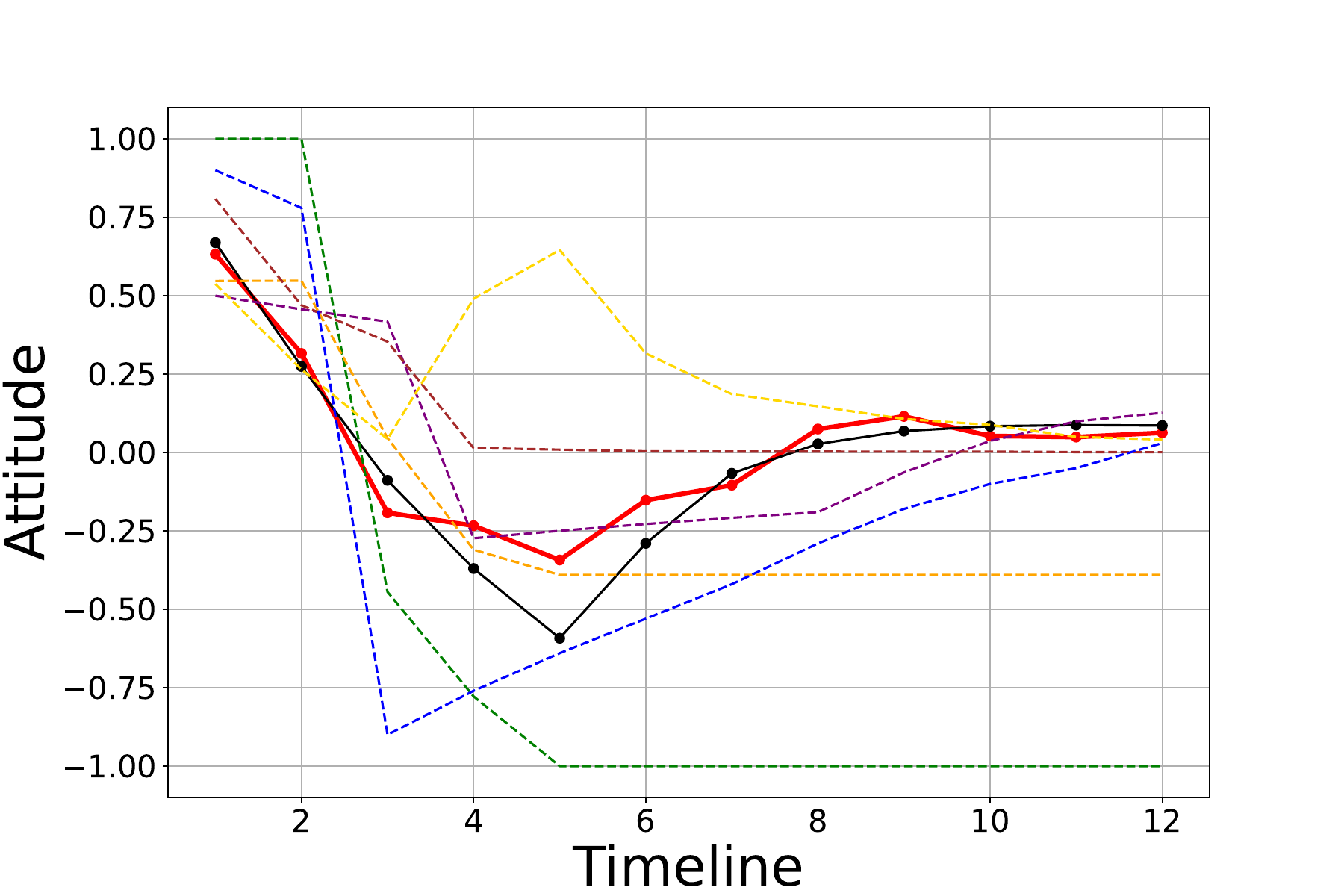}}
    % \hspace{0.04\textwidth} % fine-tuning the space between images
    \subcaptionbox{Dianduji Insident}{%
        \includegraphics[width=0.45\textwidth]{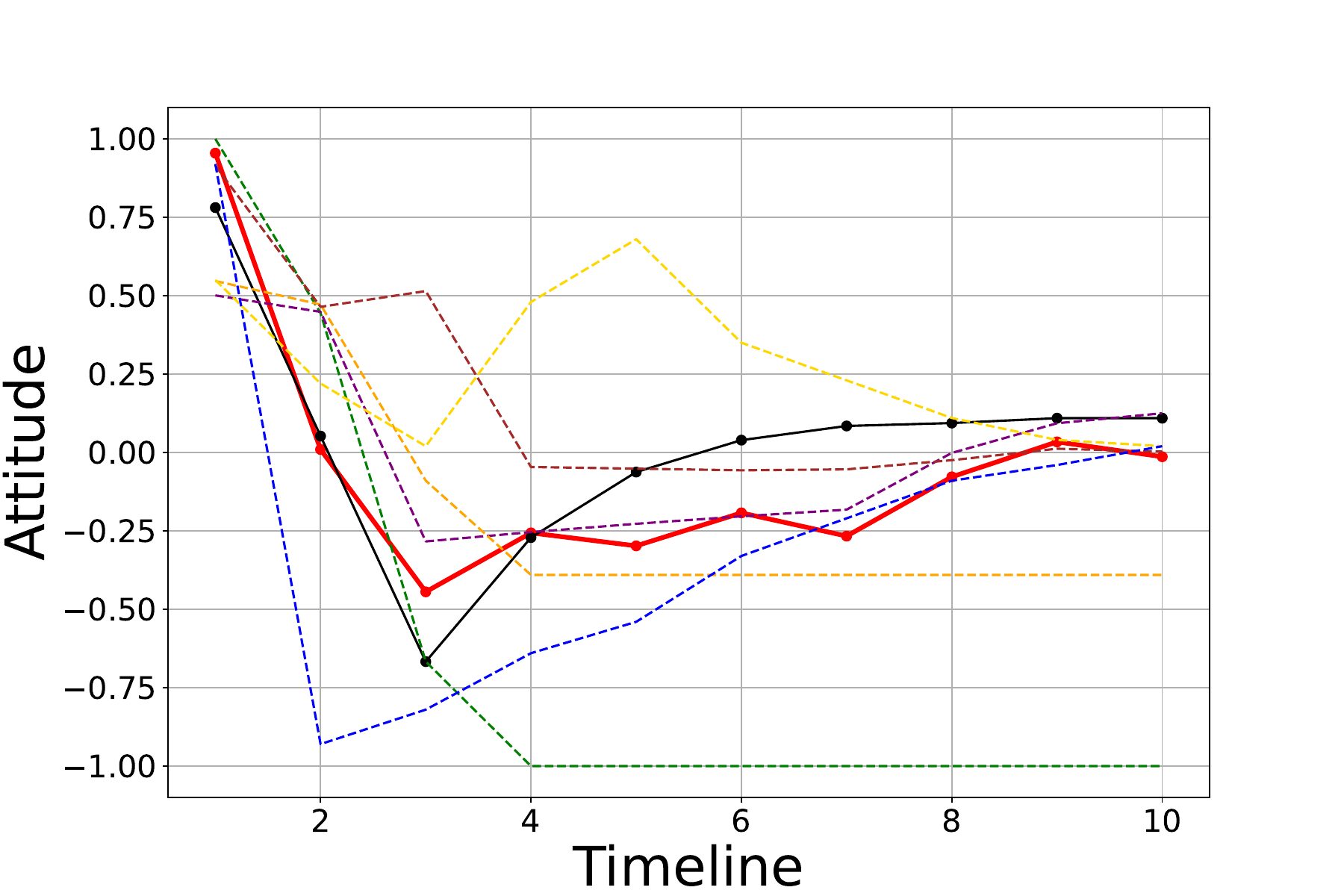}}

      \caption{Comparison of Real and Simulated Result on (a) Pangmao Incident (b) Jiangping Incident (c) Qingdao Incident (d) Dianduji Incident. The red lines represent actual data. }
           \label{comparison of real and simulated resultes}
\end{figure}
\begin{table*}[htbp]
    \centering
    \caption{Comparison of Real and Simulated Results. The best results are highlighted by \textbf{Bold} and the second-best results are marked by \underline{Underline}.}
    \label{comparison of methods table}
    \begin{tabular}{lcccccccccccc}
        \toprule
        & \multicolumn{2}{c}{\textbf{Pangmao}} & \multicolumn{2}{c}{\textbf{Jiangping}} & \multicolumn{2}{c}{\textbf{Qingdao}} &
        \multicolumn{2}{c}{\textbf{Dianduji}}\\
        \textbf{Method} & \textbf{DTW$\downarrow$} & \textbf{Corr.$\uparrow$} & 
        \textbf{DTW$\downarrow$} & \textbf{Corr.$\uparrow$} & \textbf{DTW$\downarrow$} & \textbf{Corr.$\uparrow$} & \textbf{DTW$\downarrow$} & \textbf{Corr.$\uparrow$}\\
        \midrule
        FDE-LLM & \textbf{0.3622} & \textbf{0.9653} & \textbf{0.3664} & \underline{0.8950} & \underline{0.3352} & \textbf{0.9605} & \textbf{0.3404} & \textbf{0.8842} \\
        LLM & 2.6584 & 0.7139 & 2.9963 & 0.7208 & 2.7316 & 0.7511 & 2.3170 & 0.7819 \\
        ABM(CA) & \underline{0.4335} & 0.8920 & 0.7648 & 0.6748 & \textbf{0.2904} & 0.6800 & \underline{0.6806} & 0.8282 \\
        ABM(HK) & 1.1838 & 0.6475 & \underline{0.6085} & 0.6961 & 1.1470 & 0.6874 & 0.9453 & 0.6760 \\
        ABM(PM) & 2.5653 & 0.1591 & 2.7977 & 0.3866 & 2.6302 & 0.0375 & 2.9916 & 0.2489 \\
        ABM(RLE) & 0.9592 & \underline{0.9489} & 1.3978 & \textbf{0.9321} & 1.8712 & \underline{0.9358} & 0.9180 & \underline{0.8304} \\
        LLM+ABM(CA) & 0.7463 & 0.5225 & 1.5748 & 0.6425 & 0.5497 & 0.9174 & 0.8684 & 0.7386 \\
        \bottomrule
    \end{tabular}
\end{table*}

This study focuses on social news events characterized by a rapid rise in attention and a quick loss of popularity. The loss of popularity directly reduces extreme (radical) attitudes and increases neutral and rational analysts. From our analysis of genuine attitude data, we found a key feature of social news events. After the event loses attention or the truth is exposed, attitudes stabilize, and the group attitude tends to approach 0 (group neutrality).

Our FDE-LLM model integrates the SIR model to constrain the attitudes of opinion followers, effectively simulating the loss characteristic of group attitudes. Compared to the LLM+ABM model, which emphasizes only the attitude changes prompted by news events, the FDE-LLM model enhances the natural attenuation of group attitudes on top of attitude changes, making the simulation of social news events more realistic. For example, as shown in Table \ref{comparison of methods table}, in the Pangmao event, the FDE-LLM model achieved the lowest DTW value of 0.3622, outperforming the LLM model by 86.38\% (DTW = 2.6584) and demonstrating significantly higher correlation (Corr = 0.9653) compared to ABM(HK) (Corr = 0.6475). Similarly, in the Jiangping event, the FDE-LLM model achieved a DTW value of 0.3664, substantially lower than LLM (DTW = 2.9963), and achieved the highest correlation of 0.8950, confirming its superior performance across multiple datasets.

ABM (PM) has advantages in simulating certain scenarios; however, its adaptability is limited in the face of dynamic real-world situations. ABM (RLE) performs well in trend simulation through group polarization, but the deviation introduced by polarization cannot be ignored. 

When utilizing LLM alone, the attitude shows a sharp turn and remains stable at extreme values. While the ABM model alone produces good results in some events, it lacks the interpretability of the FDE-LLM model. Detailed results are shown in Figure \ref{comparison of real and simulated resultes} and Table \ref{comparison of methods table}.

The FDE-LLM model can directly analyze group statements on a micro level, making it more valuable in predicting and controlling public opinion situations through the action log of the LLM opinion leader part. We have discussed this issue further in the Toy Example.

\subsection*{Ablation Study}
\begin{figure}[htbp]
    \centerline{\includegraphics[width=0.6\textwidth]{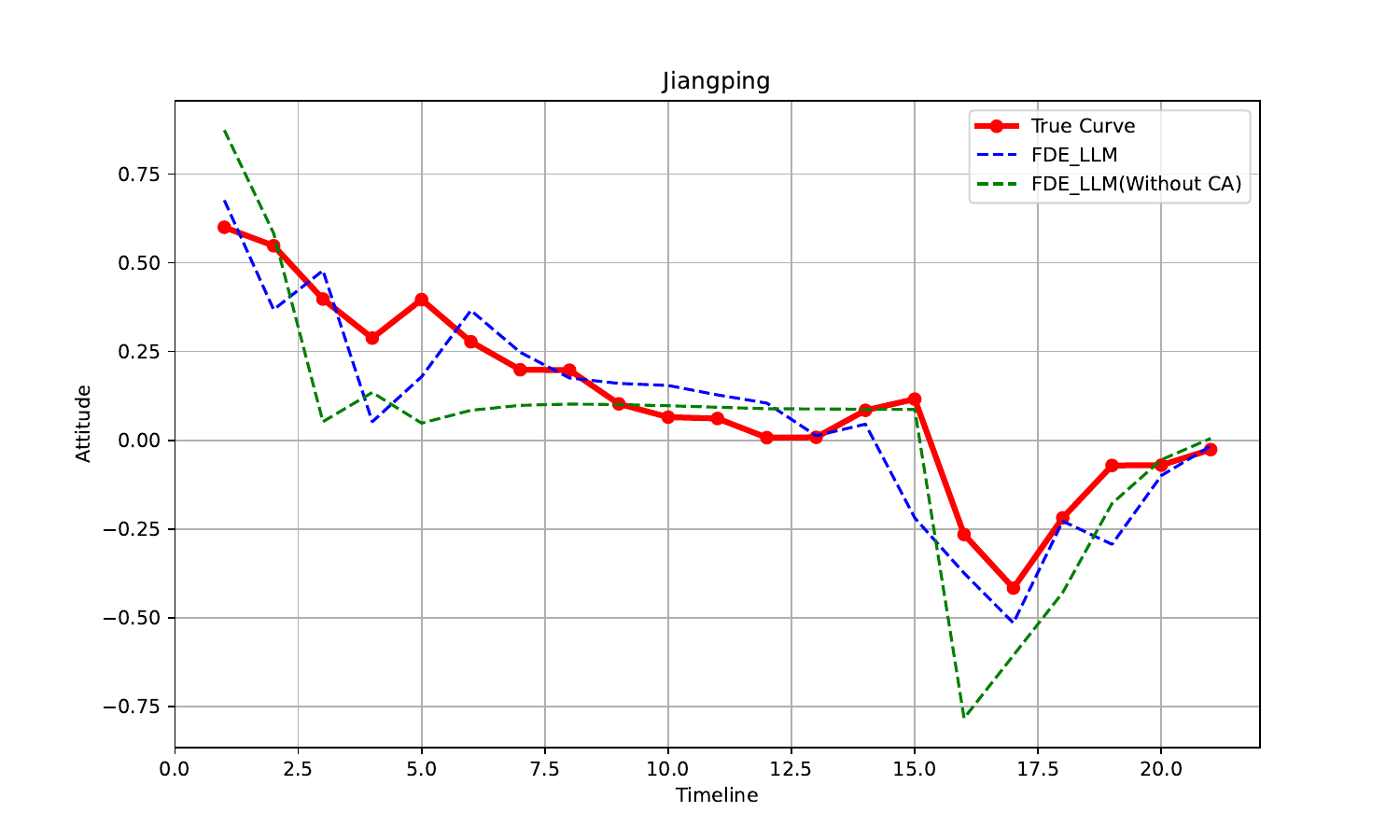}}
    \caption{FDE-LLM without CA. Without the CA constraints, the model tends to overreact in predicting extreme values, exaggerating the urgency of the situation, as observed in the Jiangping event. This overreaction can influence the decisions made by data analysts.}
    \label{fig5 ablation study}
\end{figure}

This section discusses the performance of the FDE-LLM model when the CA constraint on core users is removed. Using the Jiangping event as an example (Figure \ref{fig5 ablation study}), we observed that the model without CA constraints tends to overreact in predicting extreme values, exaggerating the urgency of the situation to some extent. For instance, the DTW value increased significantly from 0.3664 (with CA) to 0.7530 (without CA), and the correlation slightly dropped from 0.8950 to 0.8671, indicating reduced accuracy. In severe cases, this overreaction may impact the decisions made by data analysts.

The results for other datasets, as shown in Table \ref{table 2 ablation study}, reveal a similar trend, further emphasizing the importance of applying CA constraints to stabilize predictions and ensure realistic modeling of social events.
\begin{table*}[htbp]
    \centering
    \caption{Ablation result. We compared the performance of FDE-LLM with the performance after removing CA. The best data is highlighted in \textbf{Bold}. FDE-LLM outperforms the performance after removing CA in every dataset.}
    \begin{tabular}{lcccccccccccc}
        \toprule
        & \multicolumn{2}{c}{\textbf{Pangmao}} & \multicolumn{2}{c}{\textbf{Jiangping}} & \multicolumn{2}{c}{\textbf{Qingdao}} &
        \multicolumn{2}{c}{\textbf{Dianduji}}\\
        \textbf{Method} & \textbf{DTW$\downarrow$} & \textbf{Corr.$\uparrow$} & 
        \textbf{DTW$\downarrow$} & \textbf{Corr.$\uparrow$} & \textbf{DTW$\downarrow$} & \textbf{Corr.$\uparrow$} & \textbf{DTW$\downarrow$} & \textbf{Corr.$\uparrow$}\\
        \midrule
        FDE-LLM & \textbf{0.3622} & \textbf{0.9653} & \textbf{0.3664} & \textbf{0.8950} & \textbf{0.3352} & \textbf{0.9605} & \textbf{0.3404} & \textbf{0.8842} \\
        FDE-LLM (Without CA) & 0.5133 & 0.9489 & 0.7530 & 0.8671 & 0.7389 & 0.8952 & 0.5708 & 0.7367 \\
        \bottomrule
        \label{table 2 ablation study}
    \end{tabular}
\end{table*}

\begin{figure}[htbp]
    \centering
    \includegraphics[width=0.8\textwidth]{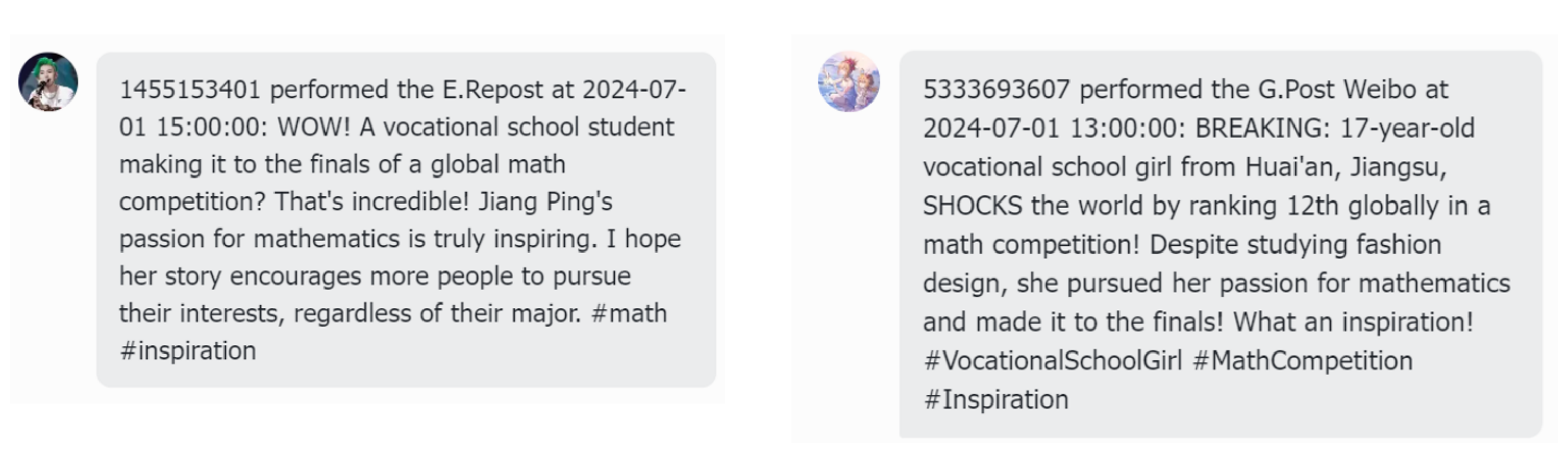}
    \caption{LLM Agent Responses Before the Reversal News}
    \label{fig6 before}
    \vspace{10pt}  % Add vertical space between the images if needed

    \includegraphics[width=0.4\textwidth]{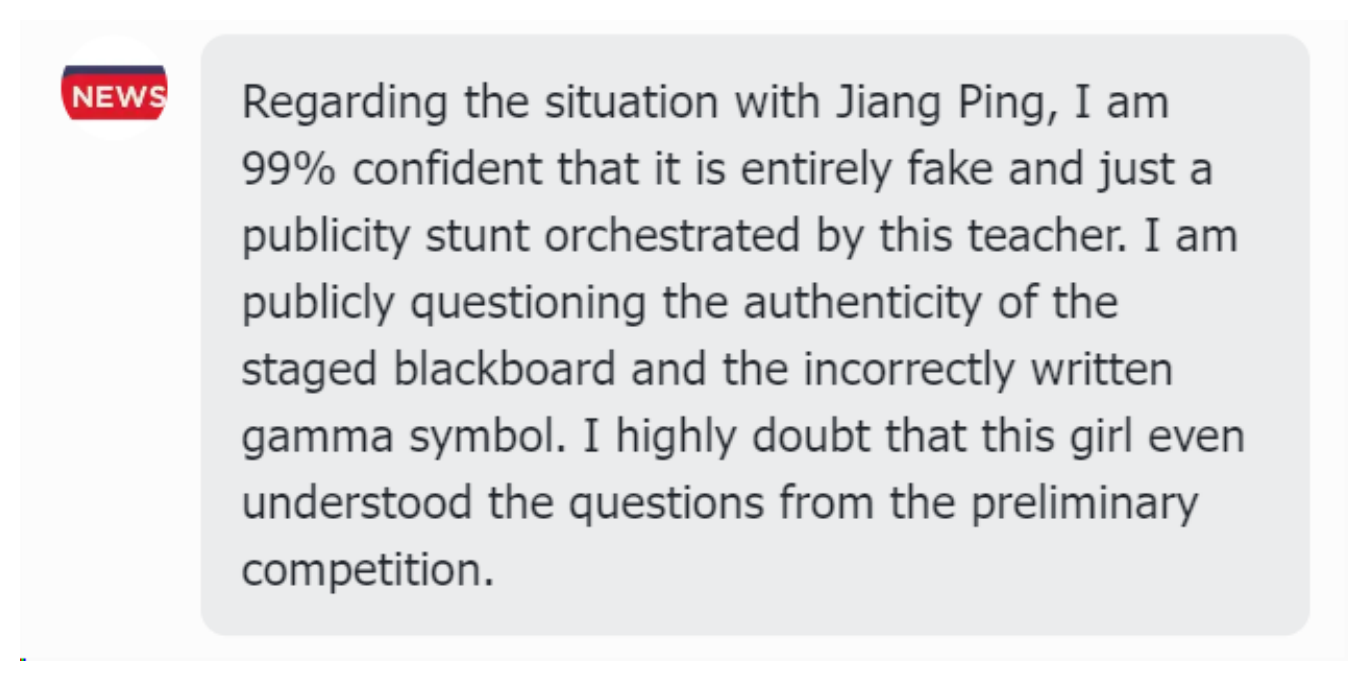}
    \caption{Reversal News}
    \label{fig7 news}
    \vspace{10pt}  % Add vertical space between the images if needed

    \includegraphics[width=0.8\textwidth]{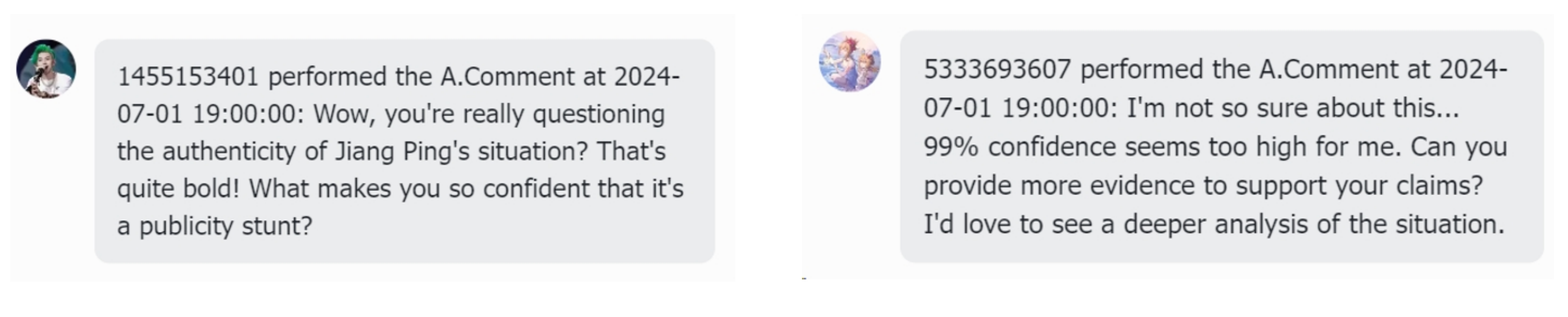}
    \caption{LLM Agent Responses After the Reversal News}
    \label{fig8 after}
\end{figure}
\subsection*{Toy Example}

We randomly selected two groups of LLM Agents to analyze their behavior before and after the release of significant news developments. In the case of the Jiangping event, before the questioning news, the Agents' attitudes were all positive (Fig \ref{fig6 before}). They expressed encouragement and anticipation and felt happy about her achievements. This response closely aligns with the actual content of Weibo posts.

However,  after the questioning news broke (Figure \ref{fig7 news}), some Agents showed strong trust in Jiangping, as shown on the left, while others agreed with the doubts, as shown on the right, which was highly consistent with the actual content of Weibo posts (Figure \ref{fig8 after}).

\section*{Conclusion}
This study optimized the model for predicting opinions by combining LLM and ABM. This approach incorporates constraints from the CA and SIR models to preserve the natural attenuation of group attitudes under the LLM guidance. Our proposed FDE-LLM categorizes users into opinion leaders and followers. Opinion leaders are stimulated using the LLM and constrained by the CA model, while opinion followers are part of a dynamic system that integrates the CA and SIR models. This innovative design significantly improves simulation accuracy and predictive efficiency, especially on reversal news. Our experimental results consistently outperform previous state-of-the-art (SOTA) simulation methods. While traditional methods may perform comparably when real-world opinion curves are relatively simple, our approach excels in handling more complex and volatile opinion dynamics. 

Future research could advance this work by integrating real-time social media data streams to enable adaptive simulations, allowing models to dynamically respond to evolving events and improve predictive robustness. Such capabilities could enhance crisis communication strategies by providing timely insights for decision-makers or inform targeted policy interventions to mitigate social polarization.
\bibliography{sample}

\section*{Author contributions statement}
H.Z. conceived the experiment(s);  J.Y., J.O., and D.Z. conducted the experiment(s); J.Y. and H.Z. analysed the results; Z.Y. and Z.D. provided suggestions. All authors reviewed the manuscript. 

\section*{Competing interests}
The authors declare no competing interests.
\section*{Data Availability}
All data generated or analysed during this study are included in this published article and supplementary information files.
\section*{Additional information}

\textbf{Source code} is available at \href{https://www.dropbox.com/scl/fo/gw3vce6rdgcfe7hr1ujrv/ALWjC8y0Hn3E-In-bSYOnxE?rlkey=usfsxhgqfaygbucsgocewinzp&st=6nzr465l&dl=0}{Code}. \textbf{Correspondence} and requests for materials should be addressed to H.Z.

\end{document}